\documentstyle[preprint,aps,epsf]{revtex}
\tighten
\begin{document}

\draft

\title{
Three-Nucleon Force Effects in Nucleon Induced Deuteron Breakup: 
Comparison to Data (II)
}

\author{
J.~Kuro\'s-\.Zo{\l}nierczuk$^1$, H.~Wita{\l}a$^1$, 
J.~Golak$^{1,2}$, H.~Kamada$^3$, A.~Nogga$^4$, R.~Skibi\'nski$^1$, 
W.~Gl\"ockle$^2$
}
\address{$^1$M. Smoluchowski Institute of Physics, Jagiellonian University,
 Reymonta 4,  30-059 Krak\'ow, Poland}
\address{$^2$Institut f\"ur Theoretische Physik II,
         Ruhr Universit\"at Bochum, D-44780 Bochum, Germany}
\address{$^3$Department of Physics, Faculty of Engineering, Kyushu Institute
of Technology \\ 1-1 Sensucho, Tobata, Kitakyushu 804-8550, Japan}
\address{$^4$Department of Physics, University of Arizona, Tucson, Arizona, 
85721, USA
}

\date{\today}
\maketitle

\begin{abstract}
Selected Nd breakup data over a wide energy range are compared to
solutions of Faddeev
equations based on modern high precision NN interactions alone and
adding current
three-nucleon force  models. Unfortunately currently available data probe
phase
space
regions for the final three nucleon momenta which are rather insensitive
to 3NF
 effects as predicted by current models. Overall there is good to
fair agreement  between present day
theory and
experiment but
also some cases exist with striking discrepancies.
Regions in the phase space are suggested where large 3NF effects can
be expected. 
\end{abstract}

\narrowtext

\section{Introduction}
\label{secIN}

In a previous paper~\cite{kur2002}, called I in the following, we performed 
a systematic search for 3NF 
effects in the full phase space of the Nd breakup process. To that 
aim we determined the 
predictions for the five-fold differential breakup cross section 
and several analyzing powers 
based on the current high-precision NN potentials AV18~\cite{AV18}, 
CD~Bonn~\cite{CDBONN}, Nijm~I,~II 
and Nijm~93~\cite{ref3} alone. 
These predictions form a band for each of the observables as a function of 
the five variables needed
for a kinematically complete determination of the breakup process. 
These five variables define 
a  point in the phase space. 
Then we added to each of 
the five NN potentials the Tucson-Melbourne three-nucleon force (TM 3NF)~\cite{coon79,ref25} 
which, with the help of a strong 
form factor parameter, has been adjusted to the $^3$H binding energy 
separately for each NN 
force~\cite{ref13}. The predictions for the observables based on these force 
combinations form another 
band. We talk of 3NF effects if the two bands are significantly 
separated. In addition we 
regarded two special cases, the NN and 3NF combinations AV18 +
Urbana~IX~\cite{ref19} and 
CD~Bonn + TM', where TM' is a 
modified TM 3NF, which corrects a 
violation of chiral symmetry in TM~\cite{ref31,ref32}. 
All the studies have been carried through with 
fully converged solutions of the Faddeev equations  for 
four nucleon laboratory energies: 
13, 65, 135 and 200 MeV. In this manner we covered a wide range 
of energies and could 
identify the different phase space regions, 
where for each of the observables 3NF effects,  
based on the current models, can be expected. 
It is now the aim of this paper to compare our predictions 
with existing data. Unfortunately, 
in contrast to Nd elastic scattering,  where precise data are 
numerous (see references in I), the existing data base 
for the breakup process is much less numerous,  
especially at  higher energies. Unfortunately,  
as we shall see,   the phase space regions, where the current 
models predict large 3NF 
effects, have not yet been explored experimentally.

Here we can not display all the existing data. For references to older 
data (before 1980)
we refer to~\cite{ref17}. We also have to omit  a very interesting full phase space 
search~\cite{BLOM}. 
Unfortunately the access to the data is no longer possible and the 
documentation in~\cite{BLOM} is insufficient to analyse the data newly. 
At that time they were analysed 
based on pioneering calculations by Kloet and Tjon~\cite{KLOT}. 
They used very simple spin 
dependent $S$-wave forces, which 
are highly insufficient by present day standards. Moreover 
those data had a high statistical 
error. Therefore we are looking forward to the data currently 
being taken  at KVI Groningen~\cite{kistryn:01,bodek}, 
which will cover a large part of the phase space, too, and will be much more accurate.

In Section~II we present a comparison of our theoretical predictions with a 
selection of more recent 
breakup data (after 1980). Most of them have been analyzed before 
by us~\cite{ref17} choosing 
either older NN potentials (Bonn B, AV14, Paris) 
or only  one  of the modern ones.
Also the addition of 3NFs has not been performed before to such an extent as in this paper. 
The criteria for the selection of data are, 
that no averaging according to acceptances and 
angular openings have to be performed, 
well documented data are available and
the experimental errors are small.~\footnote{Because of lack of other data 
we had to include some with large error bars.} 
Further we favored  cases where the same
observables
 were measured by
different groups and we tried to cover the total phase space as much as
possible.
For other data known to us (after 1980) and not shown 
we provide at least references. We close with a brief summary in Section~III.

\section{Comparison to the  data }
\label{sec:breakup-exp}
There are obviously continuously varying breakup configurations and  the
experimental
groups had to make a choice.
Up to now  so called specific configurations like FSI, QFS, STAR, and COLL 
have mostly been  measured. Their meaning will be explained
below together with
the discussion of the data.  We have chosen data at 13~MeV
representing the low energy region and at 65~MeV for the higher energy
region.
Recently new data appeared at 200 MeV~\cite{pairsuwan:95}, 
which we will also show.

As described in the introduction our theoretical predictions will be
displayed in form
of two bands corresponding to NN forces only and adding the TM 3NF. In
addition there will
be two curves for the combinations AV18 + Urbana IX and CD Bonn + TM'.

\subsection{Energy 13~MeV}

\begin{figure}[htbp]
\leftline{\mbox{\epsfxsize=13cm \epsffile{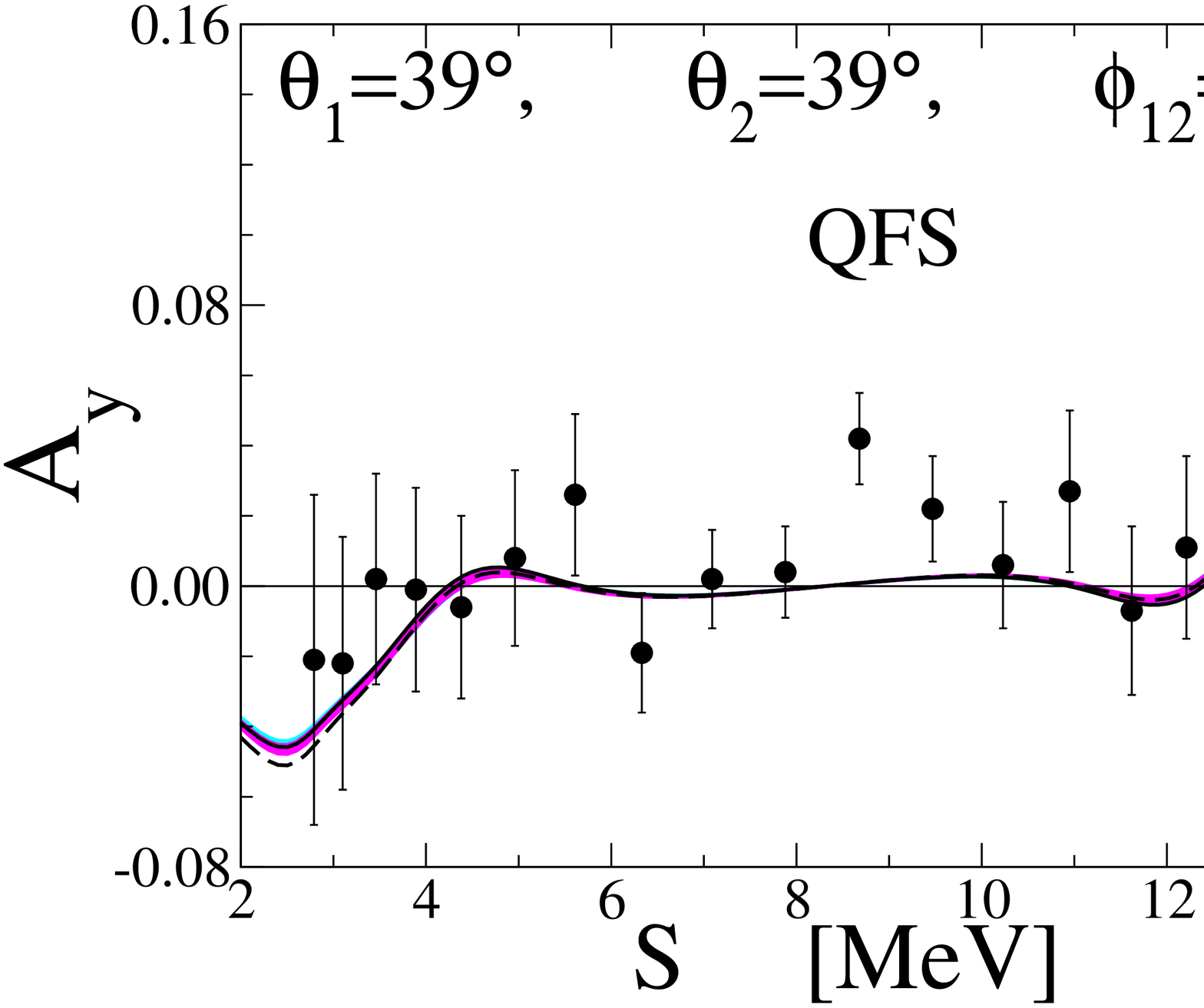}}}
\caption[]{Nd breakup cross section data in [mb MeV$^{-1}$sr$^{-2}$] 
at 13 MeV in comparison to NN force 
predictions alone (light shaded band) and adding the TM 3NF (dark shaded
band); further shown is CD Bonn + TM' (dashed line)
and AV18 + Urbana~IX (solid line). 
The  $pd$ data (full circles) are from~\cite{rauprich:91}.}
\label{fig:e13-sig-qfs}
\end{figure}

\begin{figure}[htbp]
\leftline{\mbox{\epsfxsize=13cm \epsffile{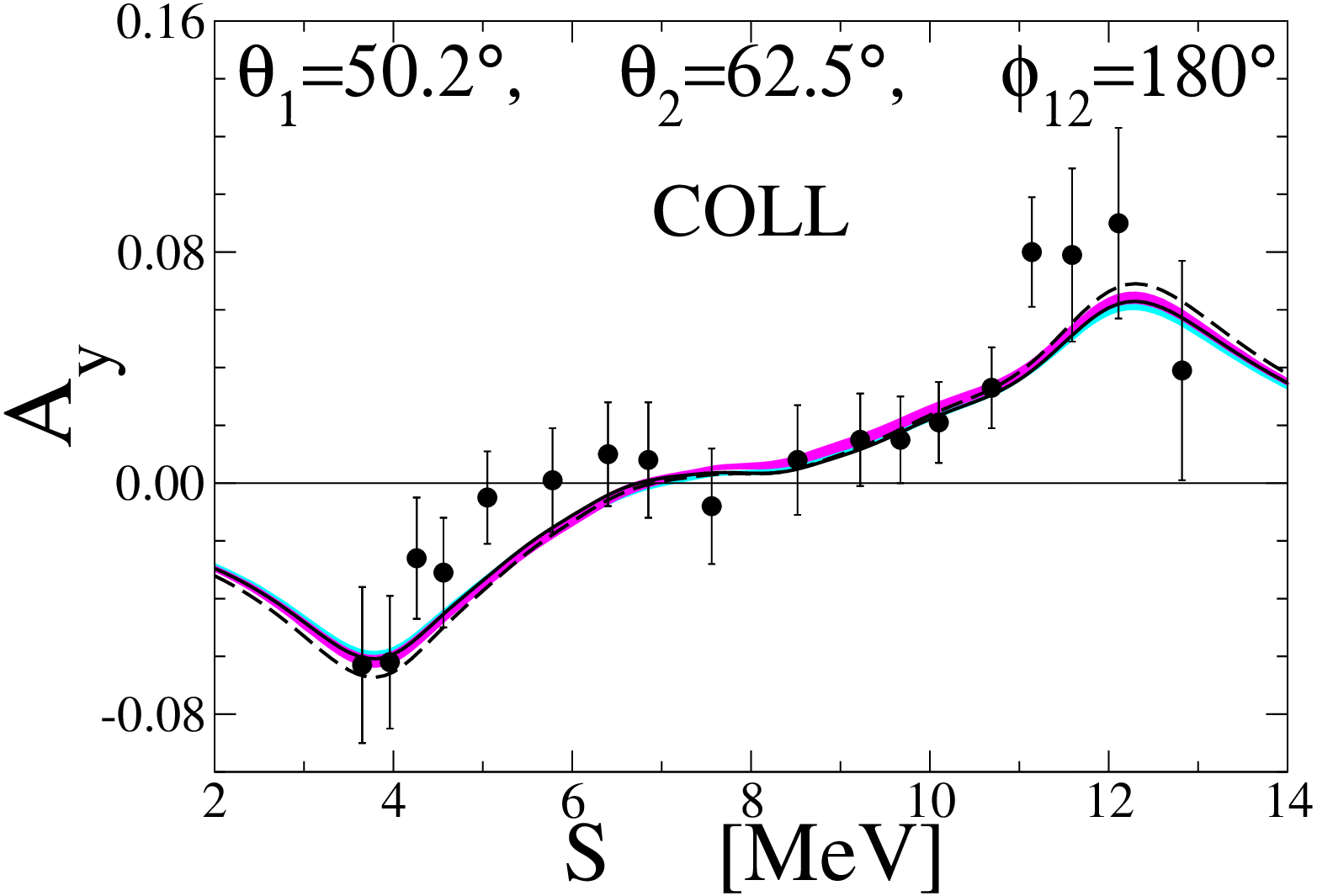}}}
\caption[]{Nucleon analyzing power $A_y$ data 
in Nd breakup at 13 MeV in comparison
to theory. Bands and curves as in Fig.~\ref{fig:e13-sig-qfs}. 
The $pd$ data are from~\cite{rauprich:91}.}
\label{fig:e13-ay-coll}
\end{figure}

The majority of the breakup experiments were performed in the region of 
low energies ($\lesssim$25~MeV) for both the 
$nd$~\cite{strate:89,stephan:89,gebhardt:93,howell:98,setze:conf95,setze:96,tor95,vwitschann,huhn00}
and the 
$pd$~\cite{correll:87,rauprich:91,patberg:96,quin:95,foroughi:80,karus:85,przyborowski:99,paetz:01,grossman:96,ZADRO} 
breakup. We compare some of the 13~MeV data  with 
our theoretical predictions for the cross section and 
 nucleon analyzing power $A_y$
in Figs.~\ref{fig:e13-sig-qfs} and \ref{fig:e13-ay-coll}. 

\begin{figure}[htbp]
\leftline{\mbox{\epsfxsize=13cm \epsffile{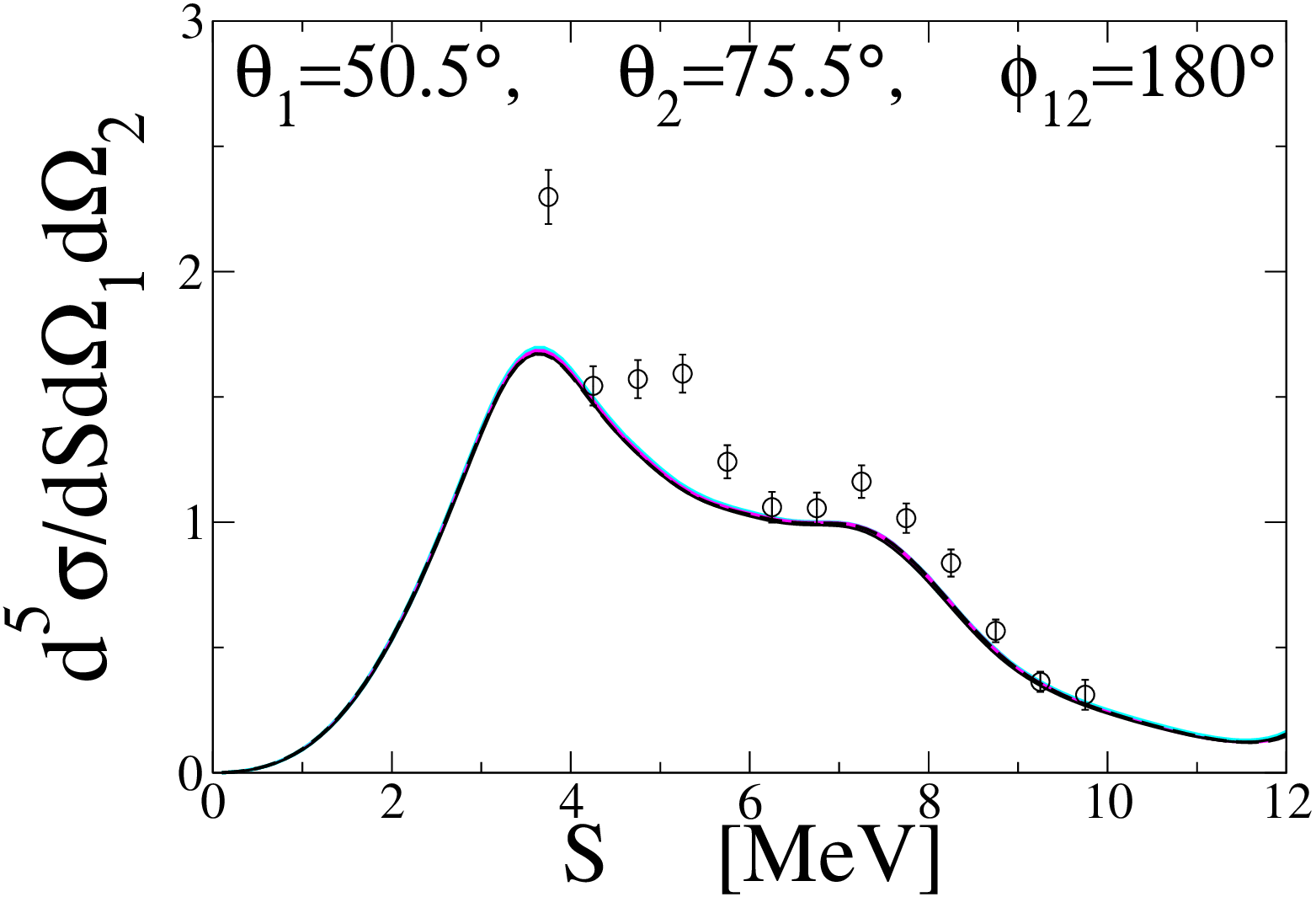}}}
\caption[]{Nd breakup cross section data in [mb MeV$^{-1}$sr$^{-2}$] 
at 13 MeV in comparison to theoretical predictions. 
Bands and curves as in~\ref{fig:e13-sig-qfs}.  
The $nd$ data are from~\cite{strate:89}.}
\label{fig:e13-sig-1}
\end{figure}

\begin{figure}[htbp]
\leftline{\mbox{\epsfxsize=13cm \epsffile{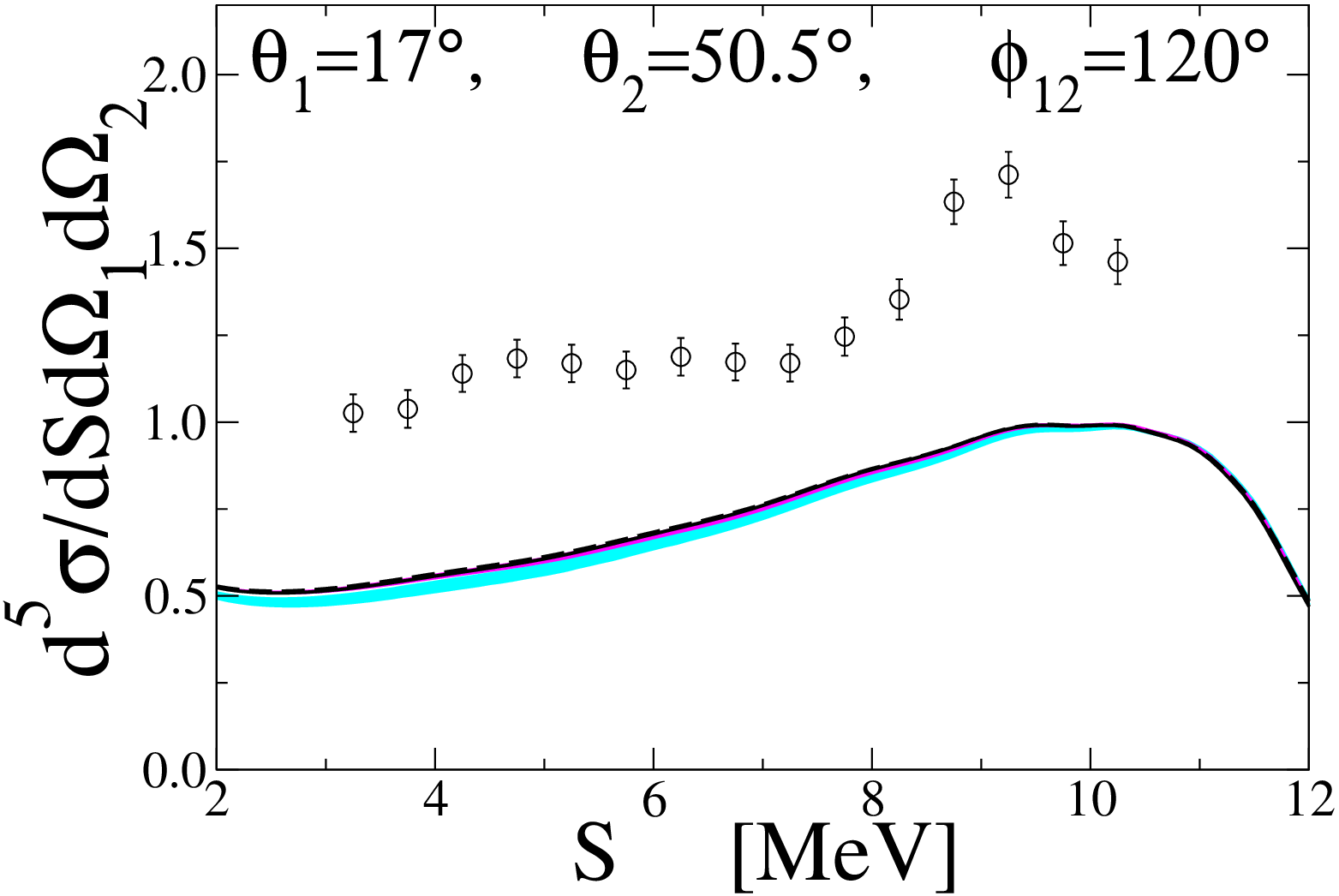}}}
\caption[]{Nd breakup cross section data in [mb MeV$^{-1}$sr$^{-2}$] 
at 13 MeV in comparison to theoretical predictions. 
Bands and curves as in~\ref{fig:e13-sig-qfs}.  
The $nd$ data are from~\cite{strate:89}.}
\label{fig:e13-sig-2}
\end{figure}

Let us first regard the cross sections which are given at the following special
configurations:
the quasi-free scattering (QFS) geometry, where one of the
nucleons in the final
state is at rest in the laboratory frame; the final state interaction
(FSI) geometry, where the relative 
energy of two outgoing nucleons is equal to
zero; the coplanar STAR geometry, where the 
three nucleons emerge from the reaction in the
c.m. system with coplanar
and equal momenta at 120$^\circ$ relative to each other and where the
beam lies
in that
plane and also the symmetric space STAR (SSS) geometry, where the c.m. plane
containing the nucleon
momenta is perpendicular to the beam direction; the collinear (COLL)
configuration, where one of
the nucleons is at rest in the c.m. system and therefore the other two
have momenta back to
back. In addition two unspecific configurations have been chosen in
Figs.~\ref{fig:e13-sig-1} and~\ref{fig:e13-sig-2}.

\begin{figure}[htbp]
\leftline{\mbox{\epsfxsize=13cm \epsffile{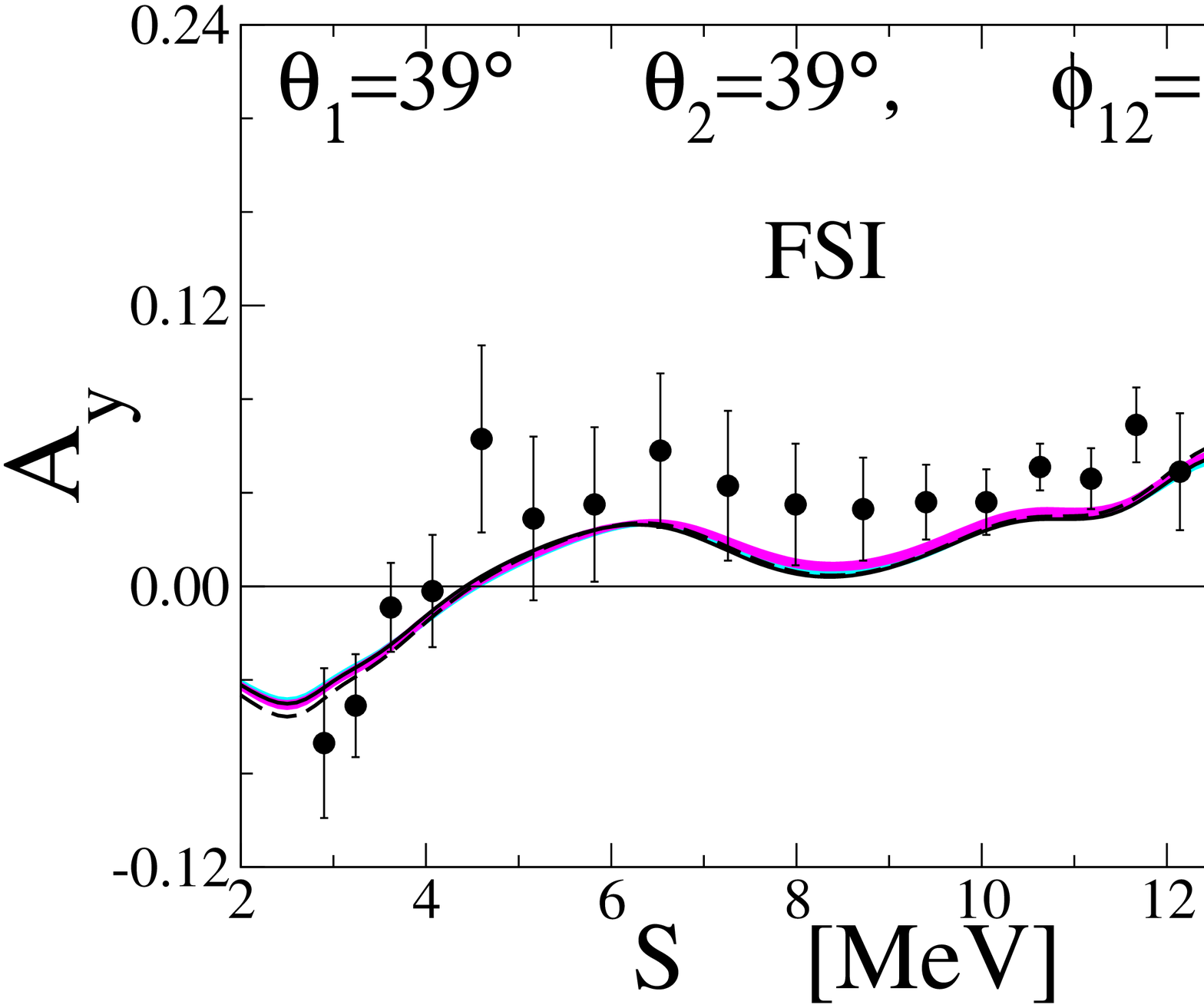}}}
\caption[]{Nd breakup cross section data in [mb MeV$^{-1}$sr$^{-2}$] 
at 13 MeV in comparison to theoretical predictions. Bands and curves as in~\ref{fig:e13-sig-qfs}.  
The $nd$ data are from~\cite{howell:98} (stars),~\cite{strate:89} (open circles),
 and the $pd$ data from~\cite{rauprich:91} (full circles).}
\label{fig:e13-sig-fsi}
\end{figure}

\begin{figure}[htbp]
\leftline{\mbox{\epsfxsize=13cm \epsffile{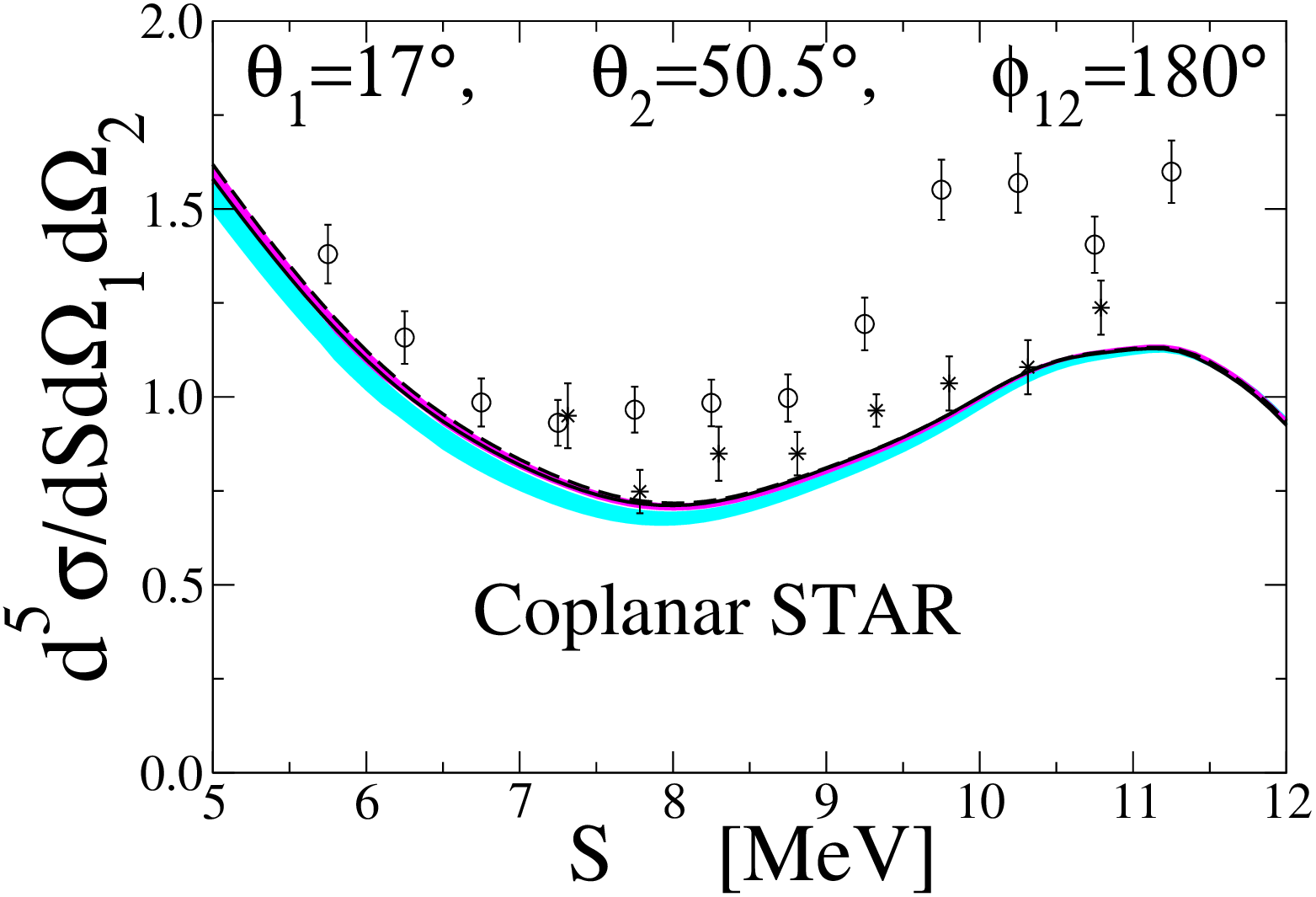}}}
\caption[]{Nd breakup cross section data in [mb MeV$^{-1}$sr$^{-2}$] 
at 13 MeV in comparison to theoretical predictions. Bands and curves as in~\ref{fig:e13-sig-qfs}.  
The $nd$ data are from~\cite{howell:98} (stars) and \cite{strate:89} (open circles).}
\label{fig:e13-sig-copstar}
\end{figure}

\begin{figure}[htbp]
\leftline{\mbox{\epsfxsize=13cm \epsffile{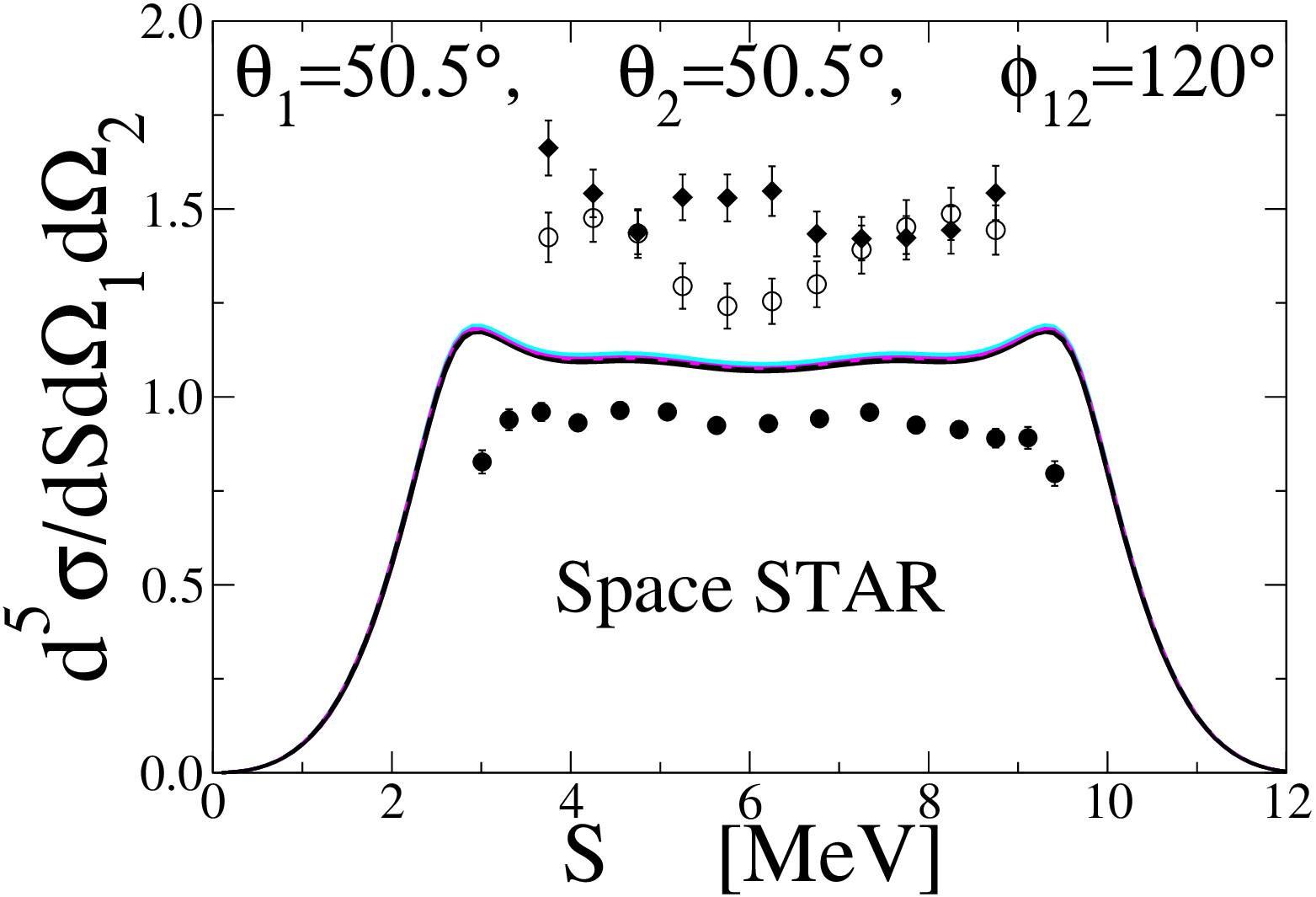}}}
\caption[]{Nd breakup cross section data in [mb MeV$^{-1}$sr$^{-2}$] 
at 13 MeV in comparison to theoretical predictions. Bands and curves as in~\ref{fig:e13-sig-qfs}.  
The $nd$ data are from~\cite{strate:89} (open circles),~\cite{setze:96} (full diamonds) and the $pd$ data from~\cite{rauprich:91} (full circles).}
\label{fig:e13-sig-spacestar}
\end{figure}

\begin{figure}[htbp]
\leftline{\mbox{\epsfxsize=13cm \epsffile{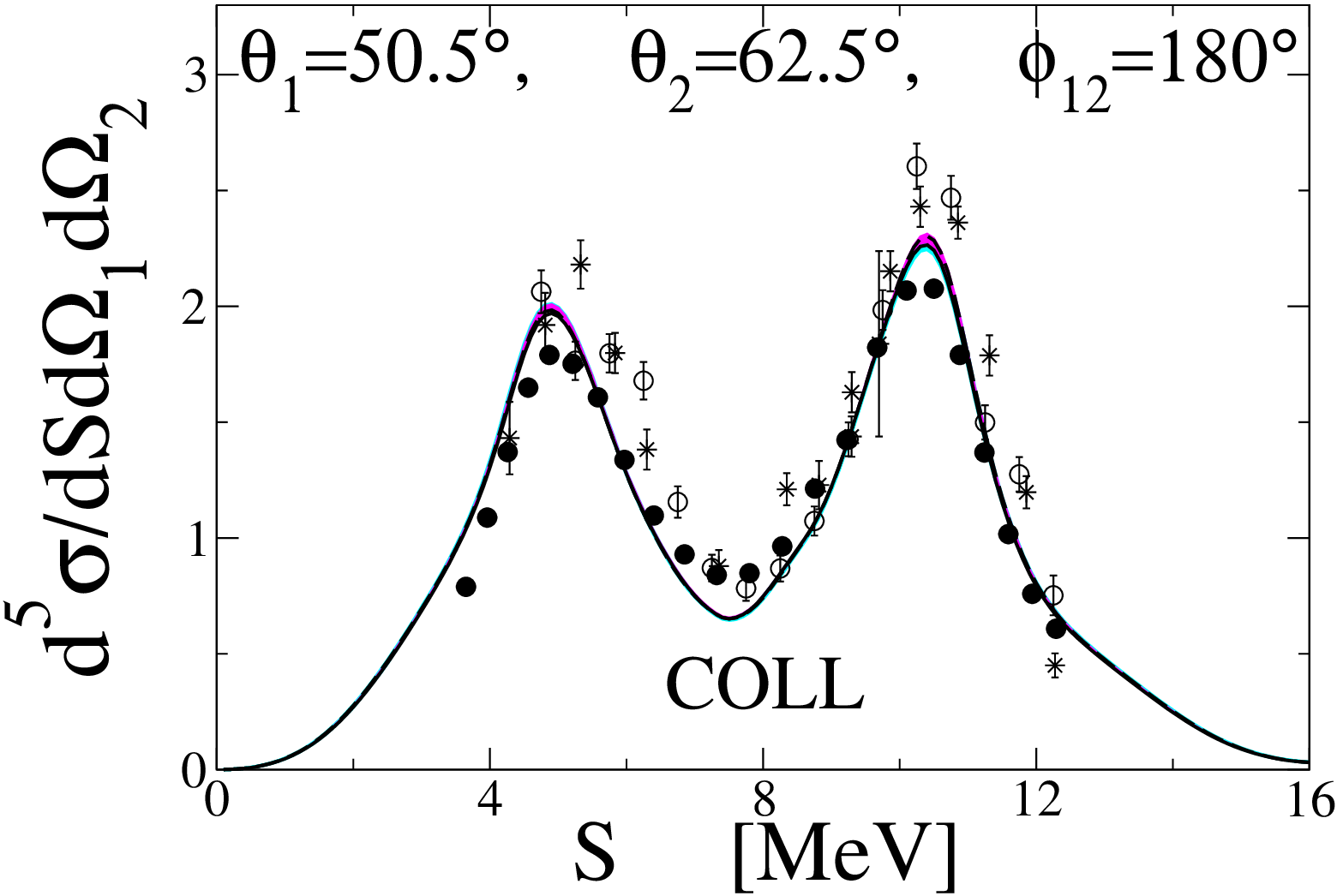}}}
\caption[]{Nd breakup cross section data in [mb MeV$^{-1}$sr$^{-2}$] 
at 13 MeV in comparison to theoretical predictions. Bands and curves as in~\ref{fig:e13-sig-qfs}.  
The $nd$ data are from~\cite{strate:89} (open circles),~\cite{strate:89} (open circles), and the $pd$ data from~\cite{rauprich:91} (full circles).}
\label{fig:e13-sig-coll1}
\end{figure}

\begin{figure}[htbp]
\leftline{\mbox{\epsfxsize=13cm \epsffile{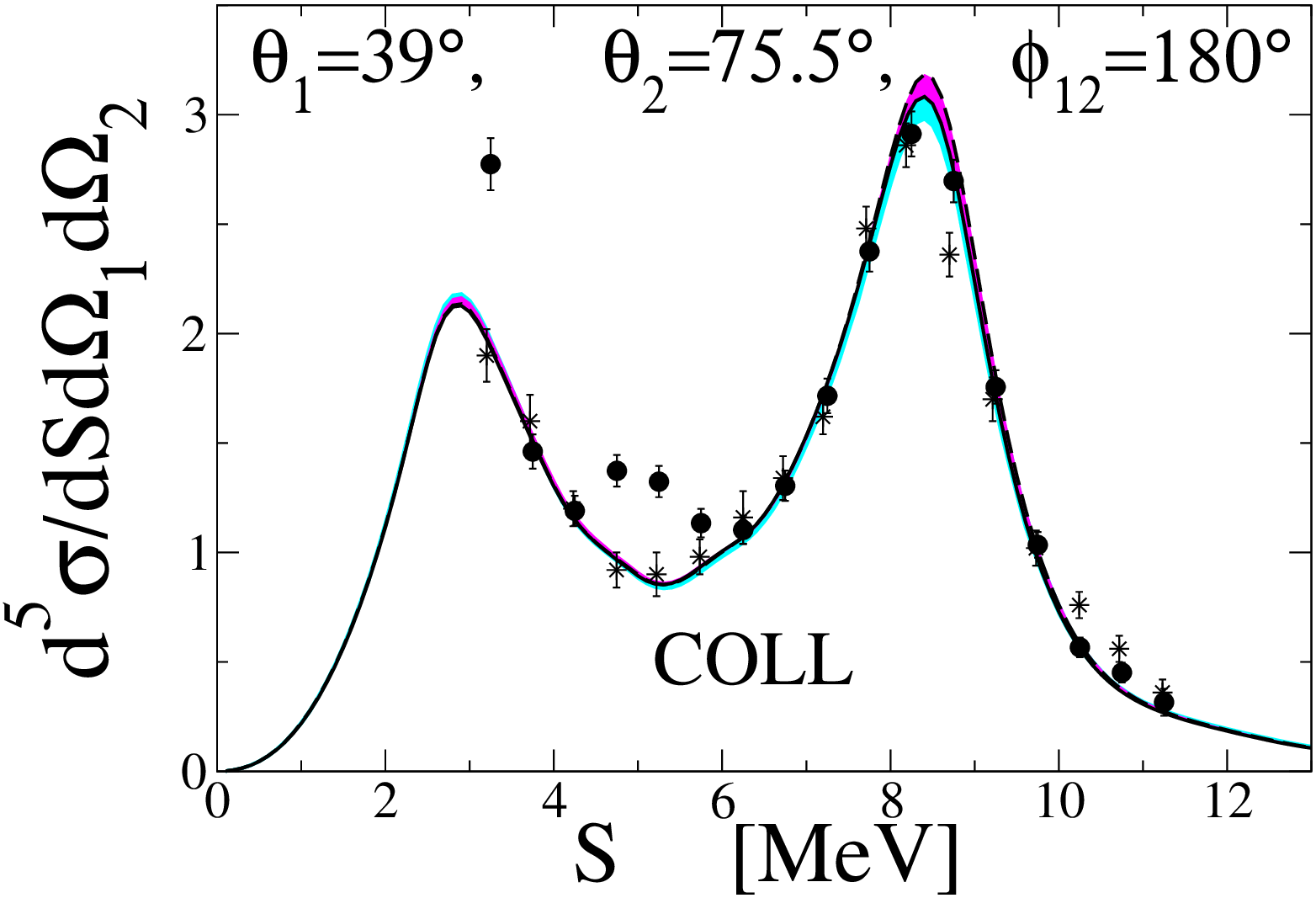}}}
\caption[]{Nd breakup cross section data in [mb MeV$^{-1}$sr$^{-2}$] 
at 13 MeV in comparison to theoretical predictions. Bands and curves as in~\ref{fig:e13-sig-qfs}.  
The $nd$ data are from~\cite{howell:98} (stars) and the $pd$ data from~\cite{rauprich:91} (full circles).}
\label{fig:e13-sig-coll2}
\end{figure}

As is seen in Figs.~\ref{fig:e13-sig-qfs}-\ref{fig:e13-sig-2}  the two bands 
are only slightly shifted to each other
and therefore 3NF
effects are very small at this energy. The pure 2N force predictions
agree in many cases
with the data. 

Especially interesting is the SSS 
configuration for which $pd$~\cite{rauprich:91} 
as well as $nd$ data taken by different 
groups~\cite{strate:89,howell:98,setze:96} exist. 
For this  configuration our theoretical $nd$ predictions shown in 
Fig.~\ref{fig:e13-sig-spacestar} 
underestimate 
the $nd$ data by  about 20\%  and overestimate the $pd$ data by about  15\%.
The discrepancy for the $pd$ data could probably have its origin in 
the neglected  $pp$ Coulomb force. 
The origin of the difference to the $nd$ data, 
called the \emph{space star anomaly}~\cite{setze:96}, is  still unknown.
The disagreement here  is quite surprising, since the calculations~\cite{setze:conf95} show that 
the NN $S$-wave contributions are the dominant part in the space star geometry 
\footnote{60\% of the space star cross section is due to the $^3S_1$ NN force, 30\%  
due to the $^1S_0$ force and only about 10\%  comes from  the $P$-wave forces~\cite{setze:conf95}.} 
and their properties are rather well determined in the NN system.

The  example with an FSI interaction peak shown in Fig.~\ref{fig:e13-sig-fsi} 
is also very interesting. This
type of peak
can be used to extract $np$ or $nn$ scattering lengths ($a_{np}$ 
or $a_{nn}$) in the state
$^1S_0$.
In such
a manner the
well known $a_{np}$ could be extracted with the correct value using only NN
forces~\cite{vwitschann,huhn00,tunlann}. In case of
$a_{nn}$ there exists a challenging controversy, where two independent $nd$
breakup measurements
lead to quite different results~\cite{vwitschann,tunlann}. 
One ~\cite{tunlann} agrees with the usually
quoted value found in
the $\pi^{-} d$ absorption process, 
while the other one~\cite{vwitschann} is significantly smaller in magnitude. 

We also display a coplanar STAR result, where a renewed
measurement~\cite{howell:98} agrees
quite well  with present day nuclear force predictions now, while an
older one~\cite{strate:89} is far off.
A corresponding shift of data occurred also for the COLL
configuration $(\theta_1,\theta_2,\phi_{12})\equiv (39^\circ,75.5^\circ, 180^\circ)$, 
where the new data~\cite{howell:98} agree with theory in contrast to the old 
one~\cite{strate:89}. 

But there are also
discrepancies. One example of QFS condition is shown in Fig.~\ref{fig:e13-sig-qfs}.
It is unknown, whether $pp$ Coulomb
force corrections  are responsible
for those deviation. A more recent measurement~\cite{VONWITSCH} 
also shows  the discrepancy for QFS conditions.
Very remarkable is also that in one of the two unspecific configurations
(17$^\circ$, 50.5$^\circ$, 120$^\circ$)
we see a dramatic disagreement of theory and data. A remeasurement would be
highly welcome.

\begin{figure}[htbp]
\leftline{\mbox{\epsfxsize=13cm \epsffile{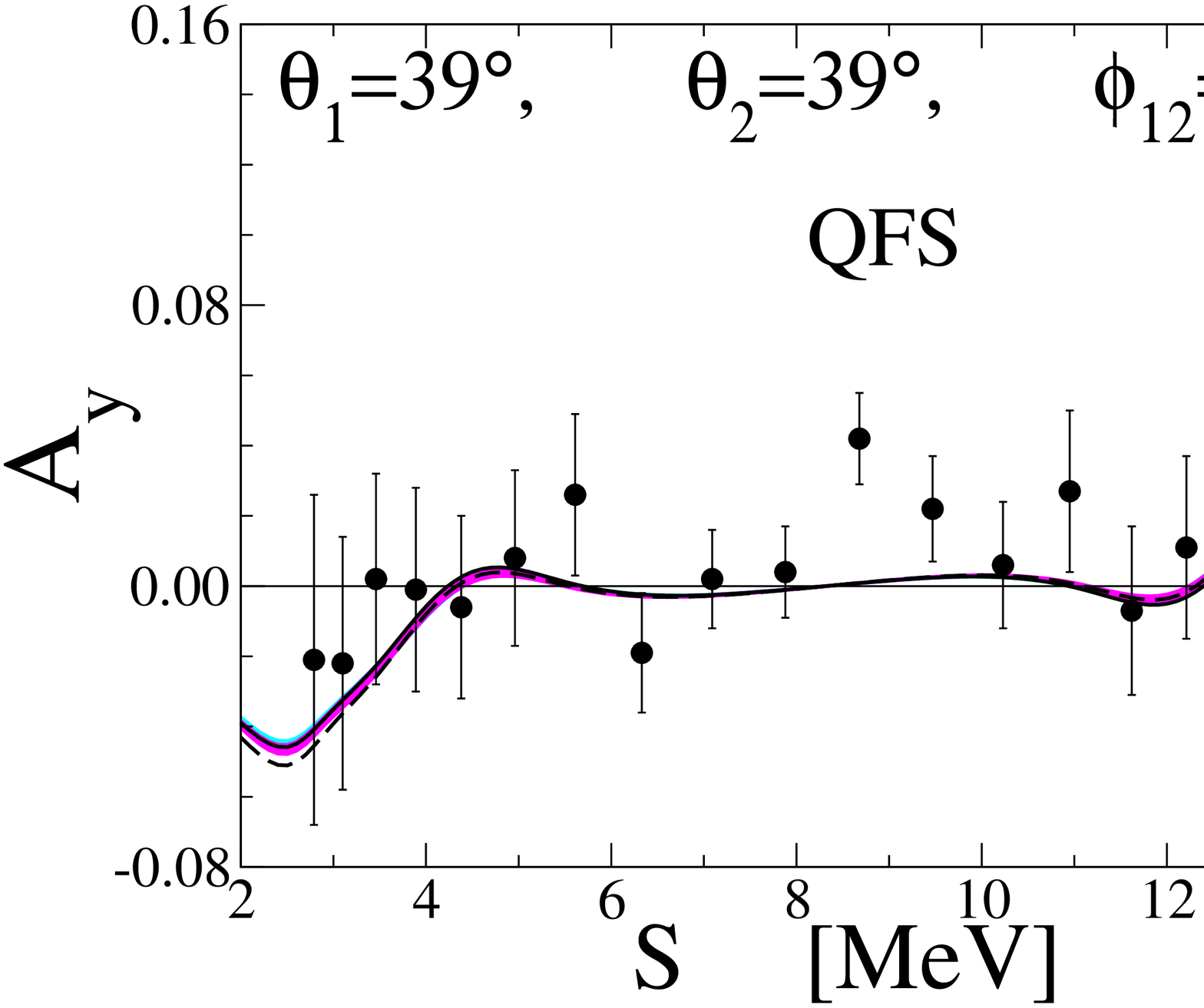}}}
\caption[]{Nucleon analyzing power $A_y$ data 
in Nd breakup at 13 MeV in comparison
to theory. Bands and curves as in Fig.~\ref{fig:e13-sig-qfs}. 
The $pd$ data are from~\cite{rauprich:91}.}
\label{fig:e13-ay-qfs}
\end{figure}

\begin{figure}[htbp]
\leftline{\mbox{\epsfxsize=13cm \epsffile{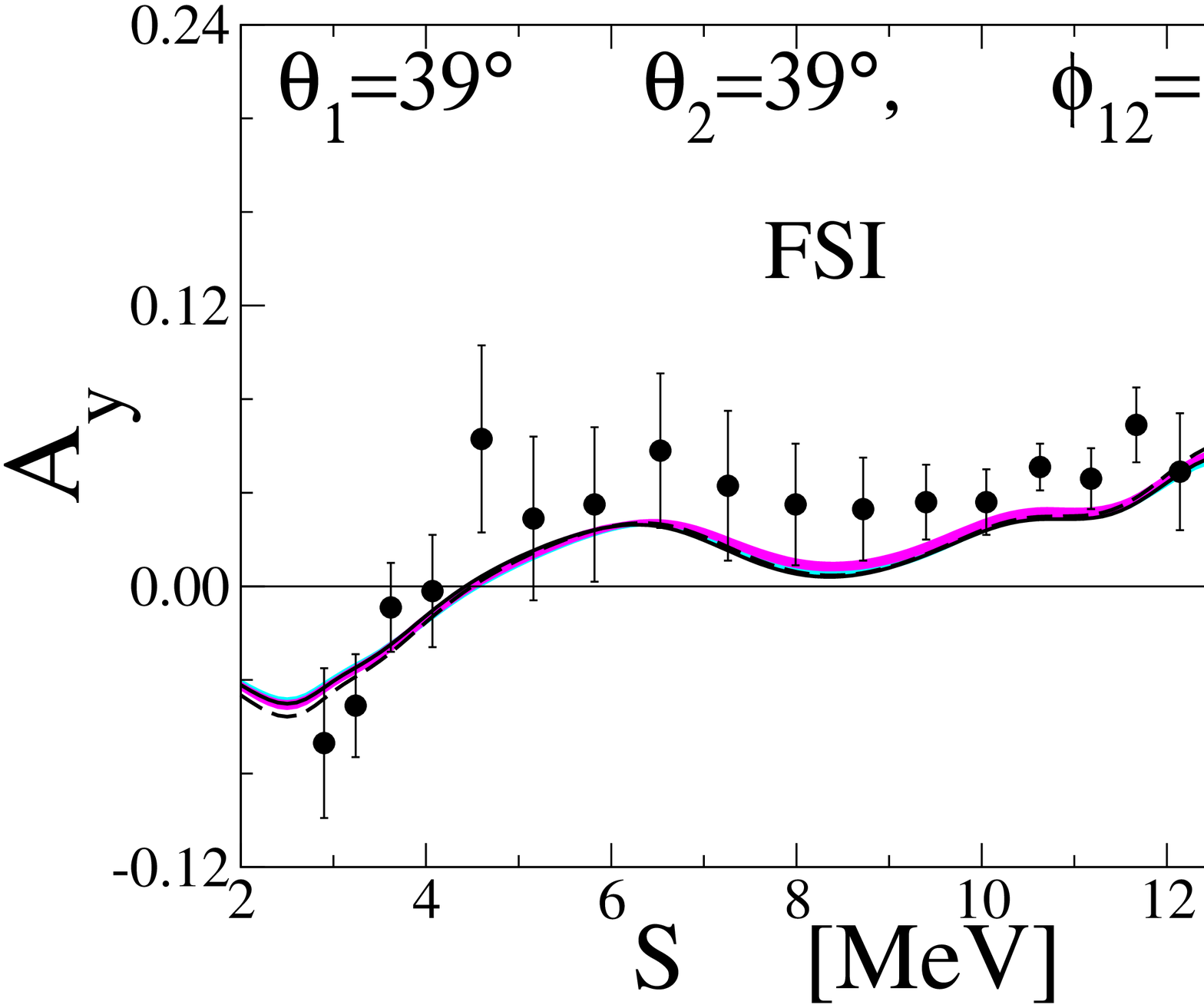}}}
\caption[]{Nucleon analyzing power $A_y$ data 
in Nd breakup at 13 MeV in comparison
to theory. Bands and curves as in Fig.~\ref{fig:e13-sig-qfs}. 
The $pd$ data are from~\cite{rauprich:91}.}
\label{fig:e13-ay-fsi}
\end{figure}

\begin{figure}[htbp]
\leftline{\mbox{\epsfxsize=13cm \epsffile{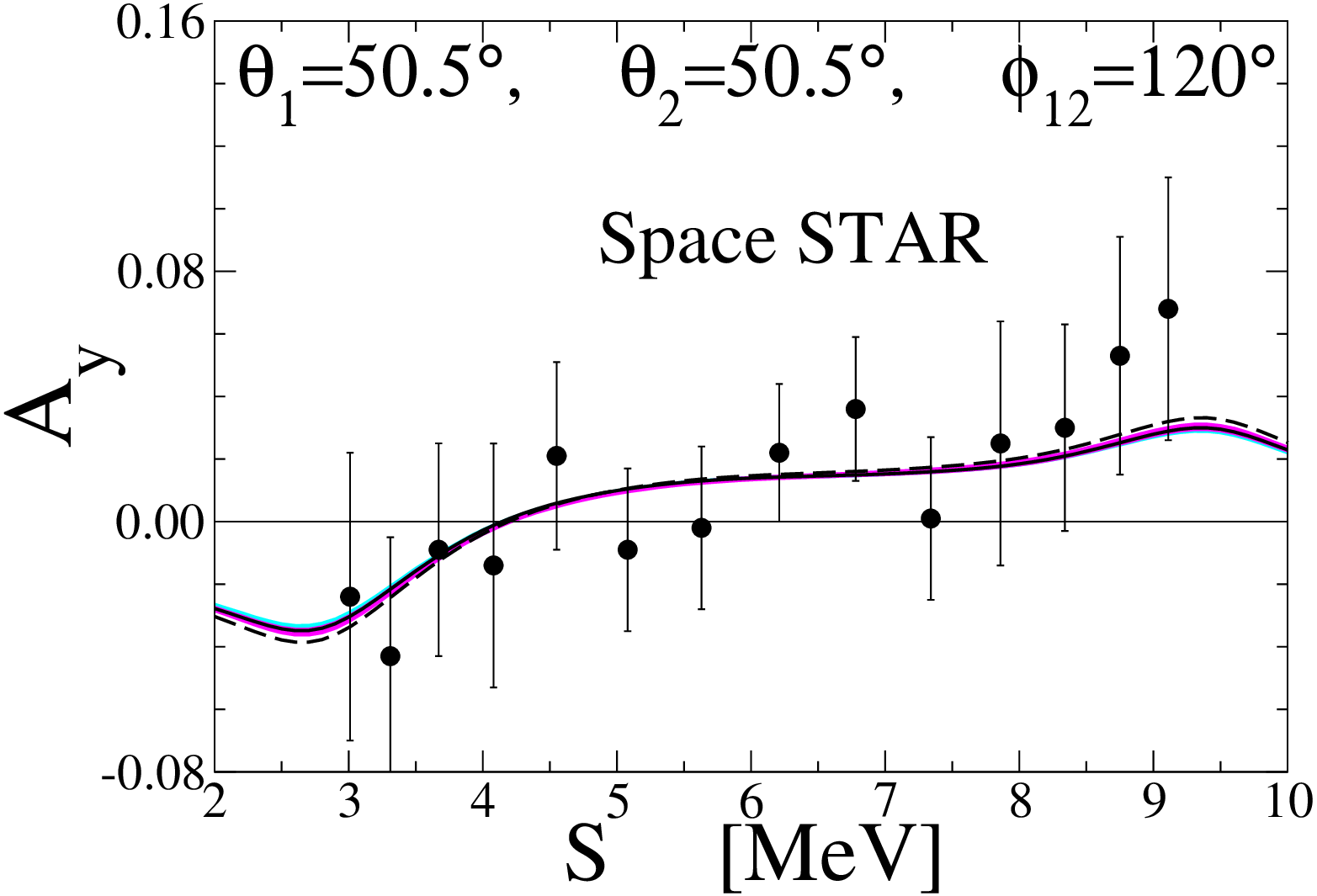}}}
\caption[]{Nucleon analyzing power $A_y$ data 
in Nd breakup at 13 MeV in comparison
to theory. Bands and curves as in Fig.~\ref{fig:e13-sig-qfs}. 
The $pd$ data are from~\cite{rauprich:91}.}
\label{fig:e13-ay-star}
\end{figure}

For the nucleon analyzing power $A_y$, 
the agreement to NN force predictions alone 
 is, in general, good (see Figs.~\ref{fig:e13-ay-qfs}-\ref{fig:e13-ay-coll}), 
though, the data scatter and have large error bars. 
All 3NFs give small effects for this observable in the chosen 
configurations at this energy.

 Further data in the low energy region can be found in~\cite{ref17}. 
The agreement
with theory is similar
as for the selected examples shown, with some further exceptions  in  the
data set from Erlangen~\cite{strate:89,gebhardt:93} and~\cite{ZADRO}.

Now, regarding the information gained in I, one has to ask whether the
available
data probed
the phase space regions, where current 3NF models predict significant
effects. The answer
is unfortunately no. For the breakup cross  section the sensitive
regions to see 3NF
effects at 13 MeV are around $\theta_1$ = $\theta_2$ = 50$^\circ$
and $\phi_{12}$= 170$^\circ$. Data  there
would be very useful. 
For the analyzing
power $A_y$ corresponding sensitive regions are around $\theta_1$ =
100$^\circ$,
$\theta_2$ = 30$^\circ$ (and vice versa) and $\phi_{12}$= 160$^\circ$.
 Unfortunately in this case 
the proton energies are rather small ($ \le $~3 MeV).

\subsection{Energy 65~MeV}

\begin{figure}[htbp]
\leftline{\mbox{\epsfxsize=17cm \epsffile{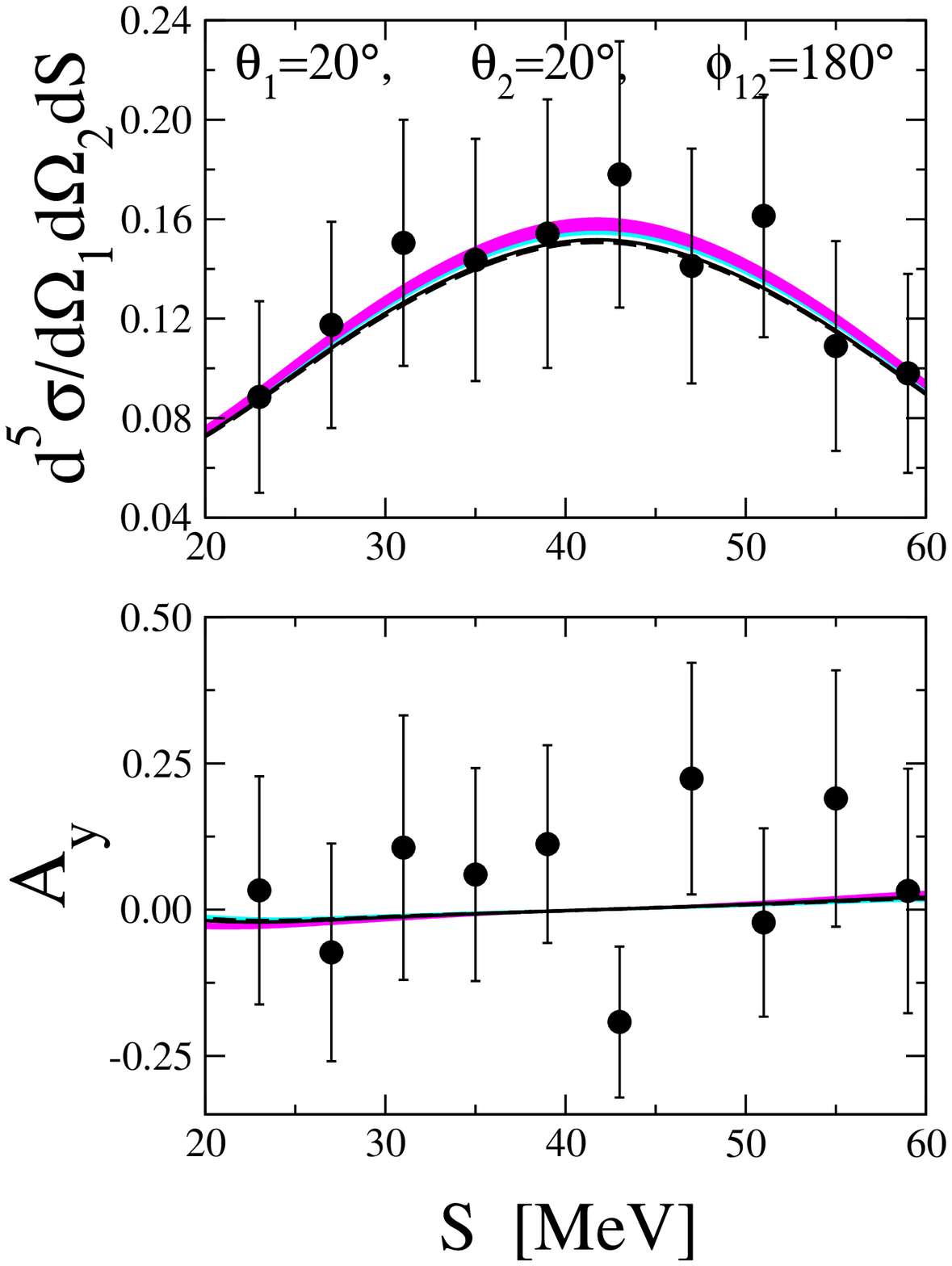}}}
\caption[]{Nd breakup cross section in [mb MeV$^{-1}$sr$^{-2}$] and nucleon analyzing power data
at 65 MeV in comparison to theory.
Unspecific configuration is shown.
Bands and curves as in Fig.~\ref{fig:e13-sig-qfs}.
The $pd$ data are from~\cite{bodek:01}.}
\label{fig:e65-1}
\end{figure}

\begin{figure}[htbp]
\leftline{\mbox{\epsfxsize=17cm \epsffile{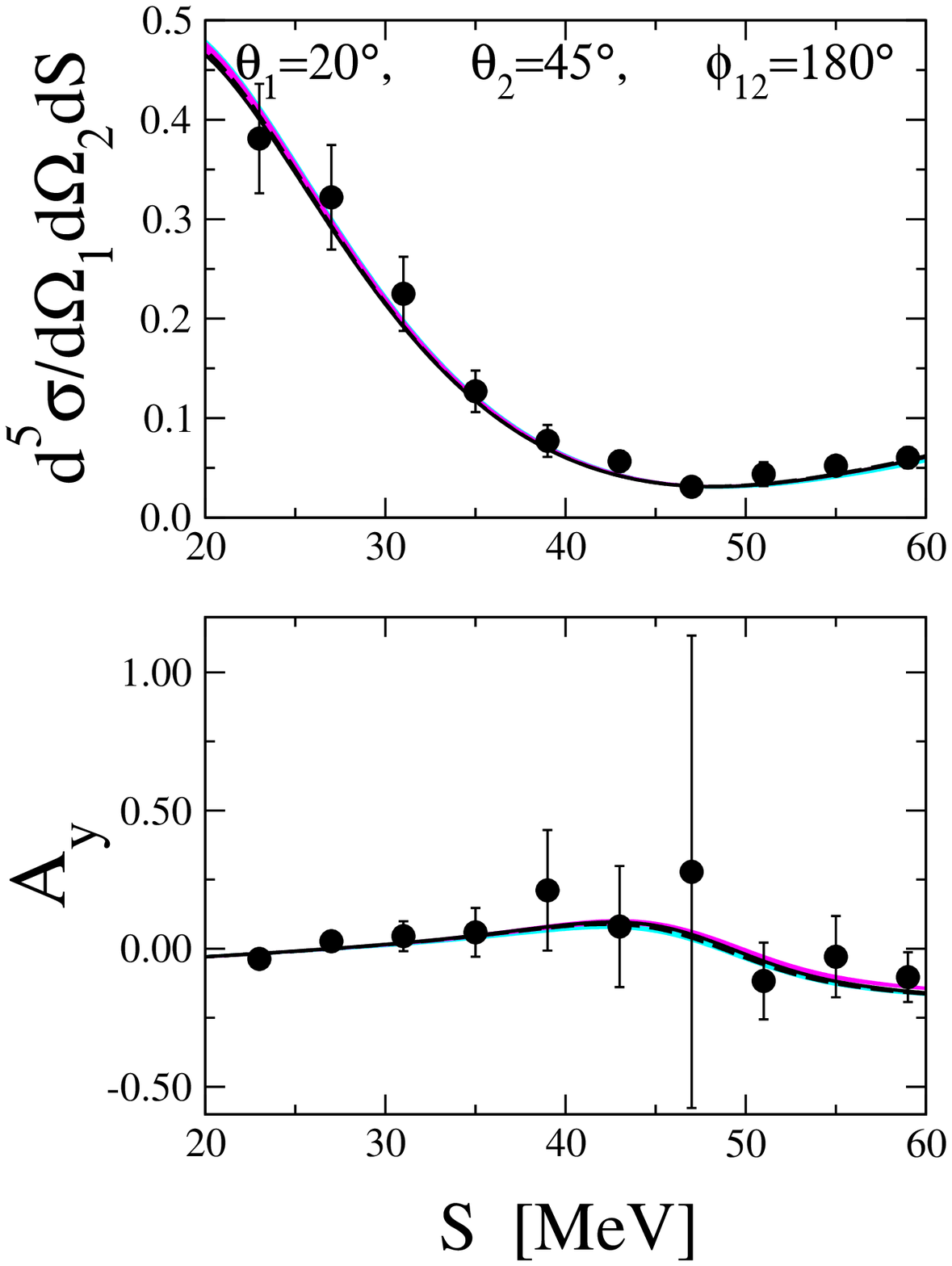}}}
\caption[]{Nd breakup cross section in [mb MeV$^{-1}$sr$^{-2}$] and nucleon analyzing power data
at 65 MeV in comparison to theory.
Unspecific configuration is shown.
Bands and curves as in Fig.~\ref{fig:e13-sig-qfs}.
The $pd$ data are from~\cite{bodek:01}.}
\label{fig:e65-2}
\end{figure}

\begin{figure}[htbp]
\leftline{\mbox{\epsfxsize=17cm \epsffile{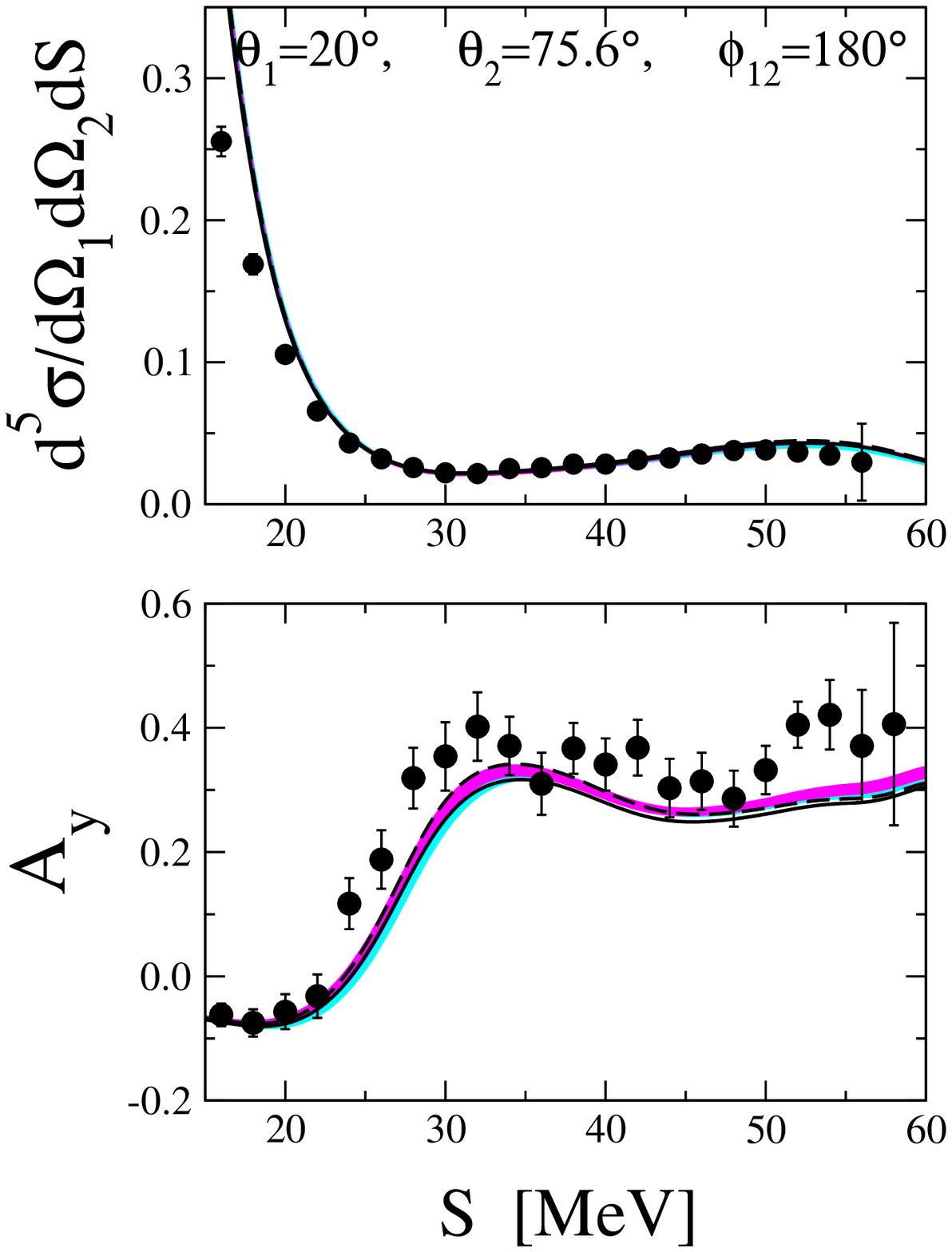}}}
\caption[]{Nd breakup cross section in [mb MeV$^{-1}$sr$^{-2}$] and nucleon analyzing power data
at 65 MeV in comparison to theory.
Unspecific configuration is shown.
Bands and curves as in Fig.~\ref{fig:e13-sig-qfs}.
The $pd$ data are from~\cite{bodek:01}.}
\label{fig:e65-3}
\end{figure}

\begin{figure}[htbp]
\leftline{\mbox{\epsfxsize=17cm \epsffile{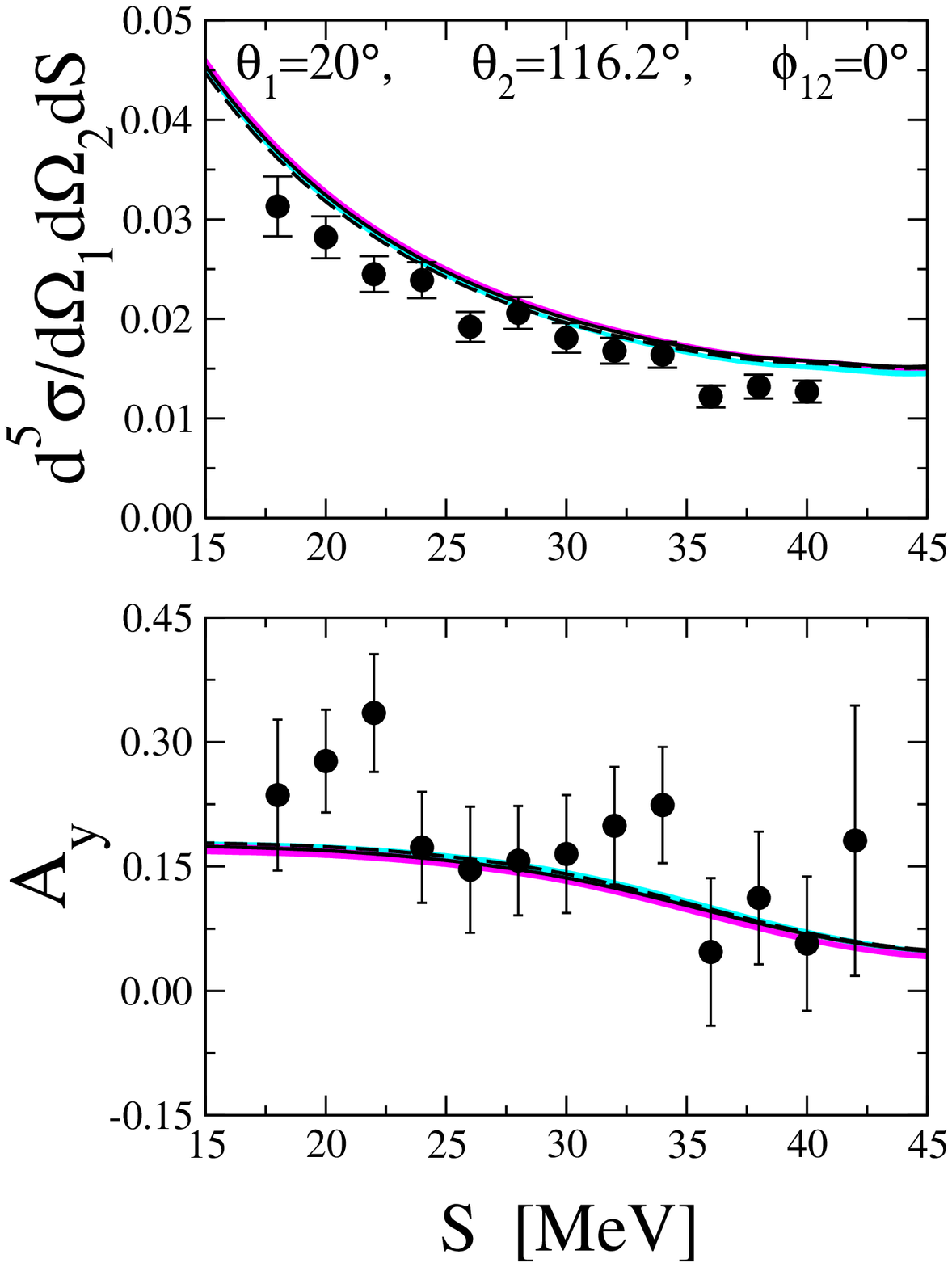}}}
\caption[]{Nd breakup cross section in [mb MeV$^{-1}$sr$^{-2}$] and nucleon analyzing power data
at 65 MeV in comparison to theory.
Unspecific configuration is shown.
Bands and curves as in Fig.~\ref{fig:e13-sig-qfs}.
The $pd$ data are from~\cite{bodek:01}.}
\label{fig:e65-4}
\end{figure}

At this energy the five-fold differential cross section and the proton  
analyzing power  were measured 
for the $\vec d(p,pp)n$ reaction  in 13 different kinematically complete  
configurations~\cite{allet:96,zejma:97,bodek:01}. 
In FIGS.~\ref{fig:e65-sss}-\ref{fig:e65-4} 
those data are compared to our theoretical predictions.

Let us first regard the cross sections. In cases where the 
two bands are narrow and either
overlap or are close together 
the agreement with the data is rather good, 
with the exception of the two 
QFS configurations (see FIGS.~\ref{fig:e65-qfs1},\ref{fig:e65-qfs2}), 
a backward plane star (BPS) configuration (see Fig.~\ref{fig:e65-bps}),  and an unspecific one
(20$^\circ$, 116.2$^\circ$, 0$^\circ$) (see Fig.~\ref{fig:e65-4}).
The BPS configuration denotes the situation where 
one of the three 
nucleons goes antiparallel to the beam direction.
 There is also forward plane star (FPS) 
configuration where one of the nucleons goes along the beam direction.
 Note in all cases one should keep 
in mind that the magnitude of the $pp$ Coulomb force effects 
under the different conditions are 
not known. For the QFS configurations one might indeed expect 
small 3NF effects, as we see, 
since by definition of that configuration one final 
nucleon is at rest and thus in a simple 
picture is like a spectator to a two-nucleon process. This 
is, however, not quite right, since 
that "spectator nucleon" is heavily rescattered as a 
comparison of the full solution with a 
plane wave assumption for that nucleon reveals~\cite{ref17,wit96}. 
Our 
results show that,  3NF effects remain 
thereby small. As we have seen at 13 MeV and what we 
found at other energies below about 
25 MeV, theory overshoots the experimental QFS 
maxima by about 20\%. This dicreases 
but remains still significant at 65 MeV with 
about 13\%. Also the QFS peak at 65 MeV 
is narrower than the theory predicts. All that might 
suggest again Coulomb force effects to
 be mostly responsible for the discrepancies. There are 
indeed first steps (based on low rank 
NN forces) which point to quite large Coulomb force effects 
for the breakup cross section~\cite{alt:94}.

In the two cases in Fig.~\ref{fig:e65-sss} and  \ref{fig:e65-coll-98}
 where the two bands are distinct (say
larger than 10\% )
the situation is controversial. 
In one case (SSS) NN predictions alone touch at least the error bars
but 3NFs move theory away from the data. In the other case, a COLL
one, neither NN forces alone nor the addition of 3NFs leads to an
agreement with the data.

Like at 13 MeV the SSS configuration poses a question.
It has been measured at several energies. 
In all cases the $pd$ data lie 
below the theoretical predictions, but this discrepancy 
decreases with increasing energy (about 15\% at 
10.5~MeV~\cite{stephan:89,gebhardt:93} and 
13~MeV and about 7\% at 19~MeV~\cite{patberg:96} and 65~MeV). Because of that 
decrease and the relative small 3NF effects one faces possibly again $pp$ 
Coulomb force effects.

For $A_y$, like for $\sigma$,  in the cases where the two bands 
are narrow and essentially overlapping there is 
agreement with the data with the exception 
of the configuration 
(59.5$^\circ$, 59.5$^\circ$, 180$^\circ$) (see Fig.~\ref{fig:e65-coll-59.5}),
where theory is partially below and partially above the data. 
When the bands 
are wider and clearly distinct unfortunately the data 
scatter a lot (see the configurations 
(30$^\circ$, 59.5$^\circ$, 180$^\circ$)-Fig.\ref{fig:e65-qfs1},
(20$^\circ$, 116.2$^\circ$, 180$^\circ$)-Fig.\ref{fig:e65-coll-116},
(30$^\circ$, 98$^\circ$, 180$^\circ$)-Fig.\ref{fig:e65-coll-98}).
There are two more cases with less narrow 
bands ((45$^\circ$, 75.6$^\circ$, 180$^\circ$)-Fig.\ref{fig:e65-coll-75.6}
and (20$^\circ$, 75.6$^\circ$, 180$^\circ$)-Fig.\ref{fig:e65-3}),
where the data 
appear to differ from theory.

Further breakup data at and around 65 MeV can be found 
in~\cite{quin:95,allet:96,zejma:97,bodek:01,all94,allet:96qfs,all96,low91}.

Again we ask, whether the sensitive regions for 3NF effects according to
I have been included in the existing data base.
Unfortunately this is again not the case. 
The sensitive regions for the cross section and $A_y$
are around $\theta_1 \sim $ 20~$^\circ, \theta_2 \sim 10~^\circ$ 
(and vice versa) 
and $ 0^\circ \le \phi_{12}  \le 60^\circ $. Though the configuration 
$(30^\circ, 98^\circ, 180^\circ)$, for instance, in case of $A_y$ 
shows an interesting sensitivity to 3NFs, the effects are only of $30\%$, 
whereas effects of up to $100\%$ and higher are predicted in the 
geometries just mentioned. 

\subsection{Energy 200~MeV}

\begin{figure}[htbp]
\leftline{\mbox{\epsfxsize=17cm \epsffile{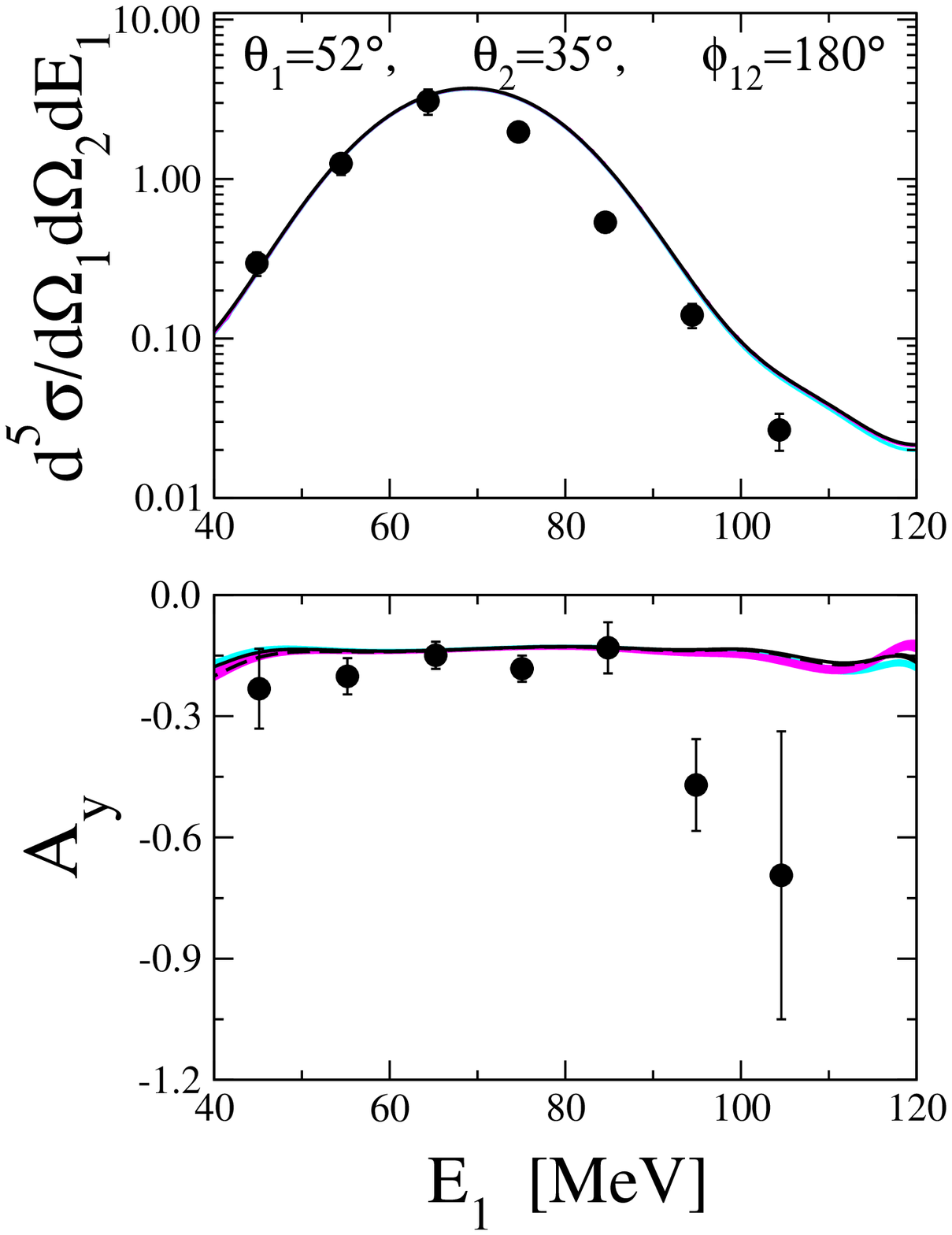}}}
\caption[]{Nd breakup cross section in [mb MeV$^{-1}$sr$^{-2}$]
and nucleon analyzing power data at 200 MeV in comparison to theory. 
Bands and curves as in Fig.~\ref{fig:e13-sig-qfs}. The $pd$ data are from~\cite{pairsuwan:95}.}
  \label{fig:e200-1}
\end{figure}

\begin{figure}[htbp]
\leftline{\mbox{\epsfxsize=17cm \epsffile{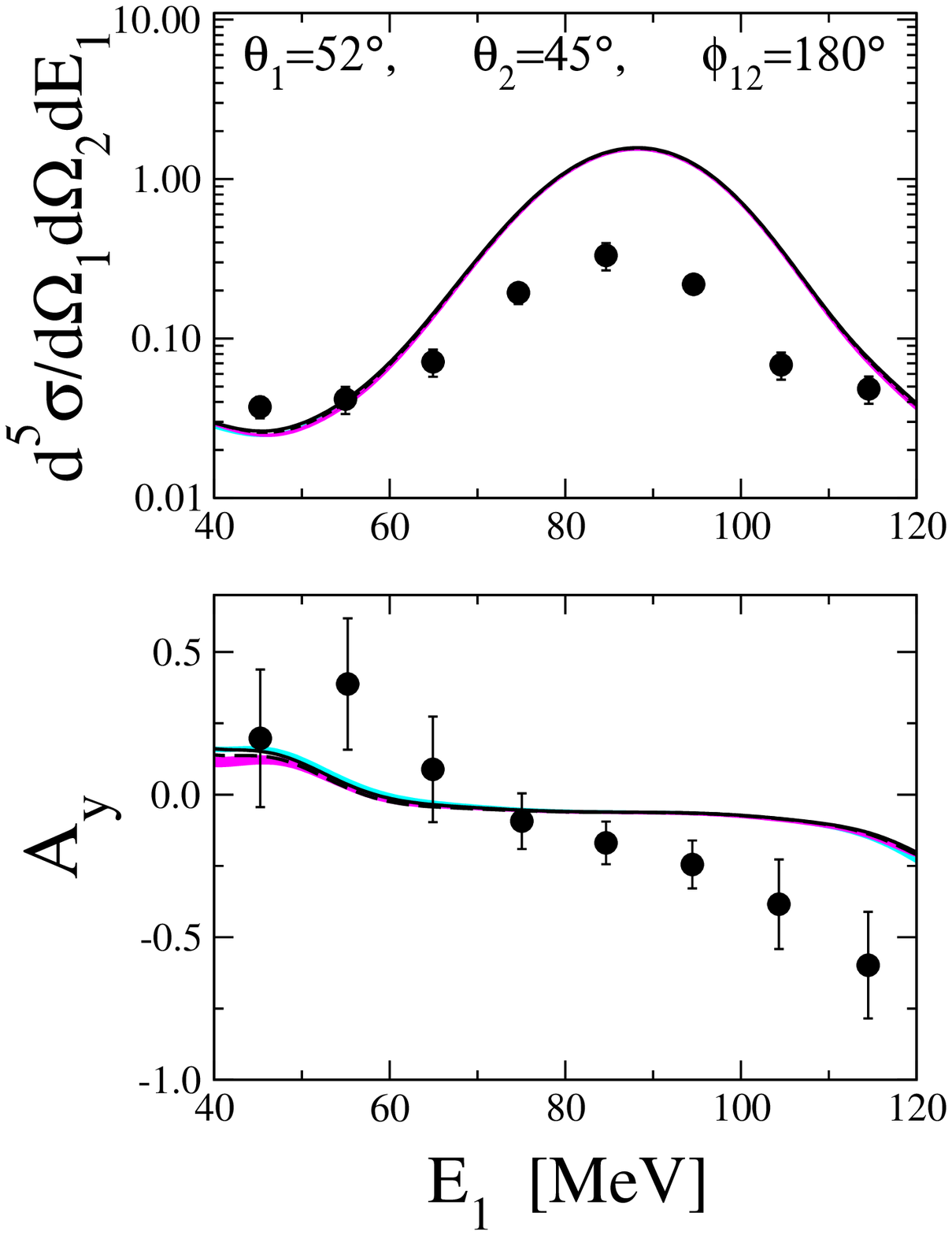}}}
  \caption[]{Nd breakup cross section in [mb MeV$^{-1}$sr$^{-2}$]
and nucleon analyzing power data at 200 MeV in comparison to theory. 
Bands and curves as in Fig.~\ref{fig:e13-sig-qfs}. The $pd$ data are from~\cite{pairsuwan:95}.
}
  \label{fig:e200-2}
\end{figure}

\begin{figure}[htbp]
\leftline{\mbox{\epsfxsize=17cm \epsffile{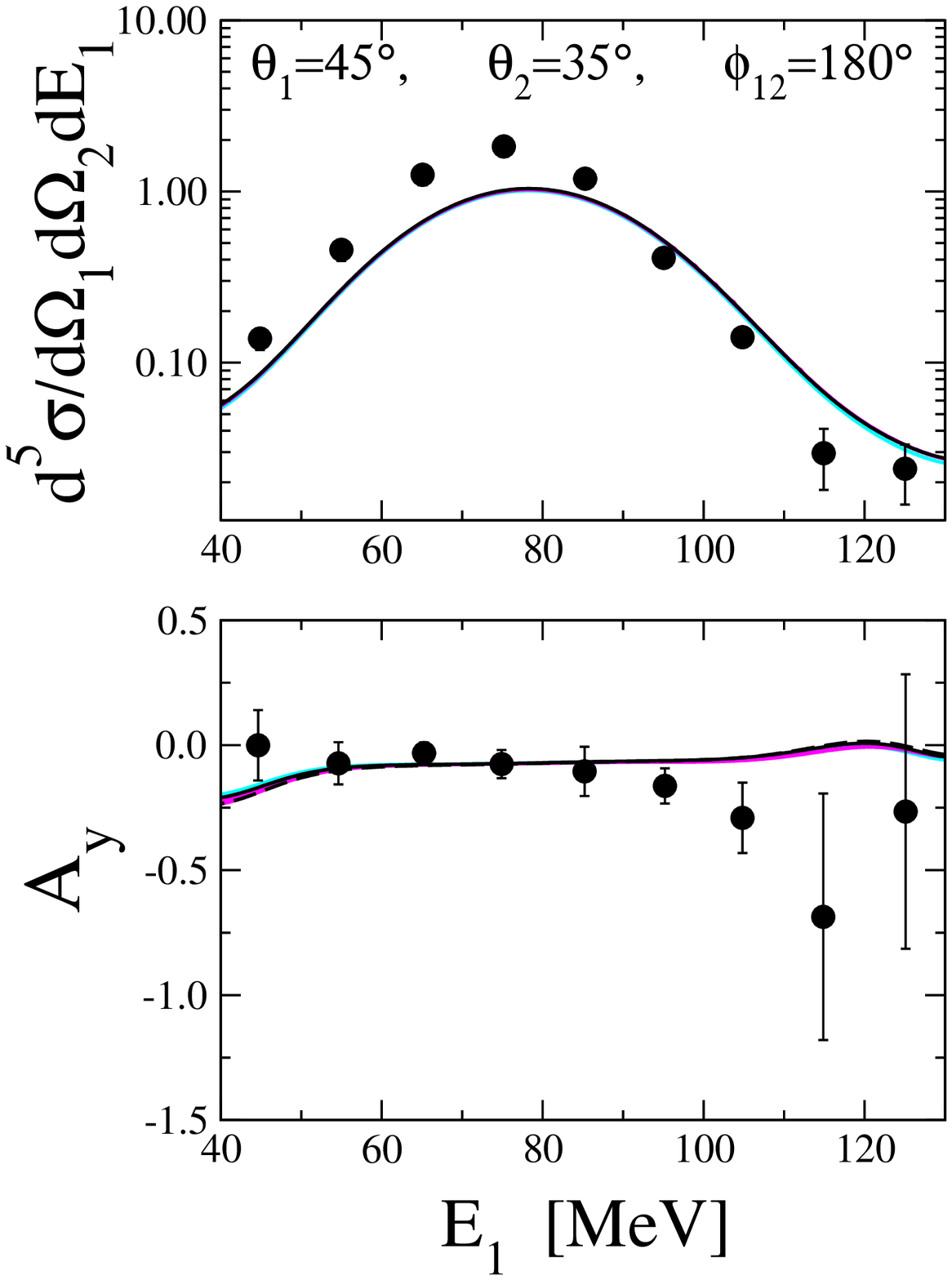}}}
  \caption[]{Nd breakup cross section in [mb MeV$^{-1}$sr$^{-2}$]
and nucleon analyzing power data at 200 MeV in comparison to theory. 
Bands and curves as in Fig.~\ref{fig:e13-sig-qfs}. The $pd$ data are from~\cite{pairsuwan:95}.
}
  \label{fig:e200-3}
\end{figure}

\begin{figure}[htbp]
\leftline{\mbox{\epsfxsize=17cm \epsffile{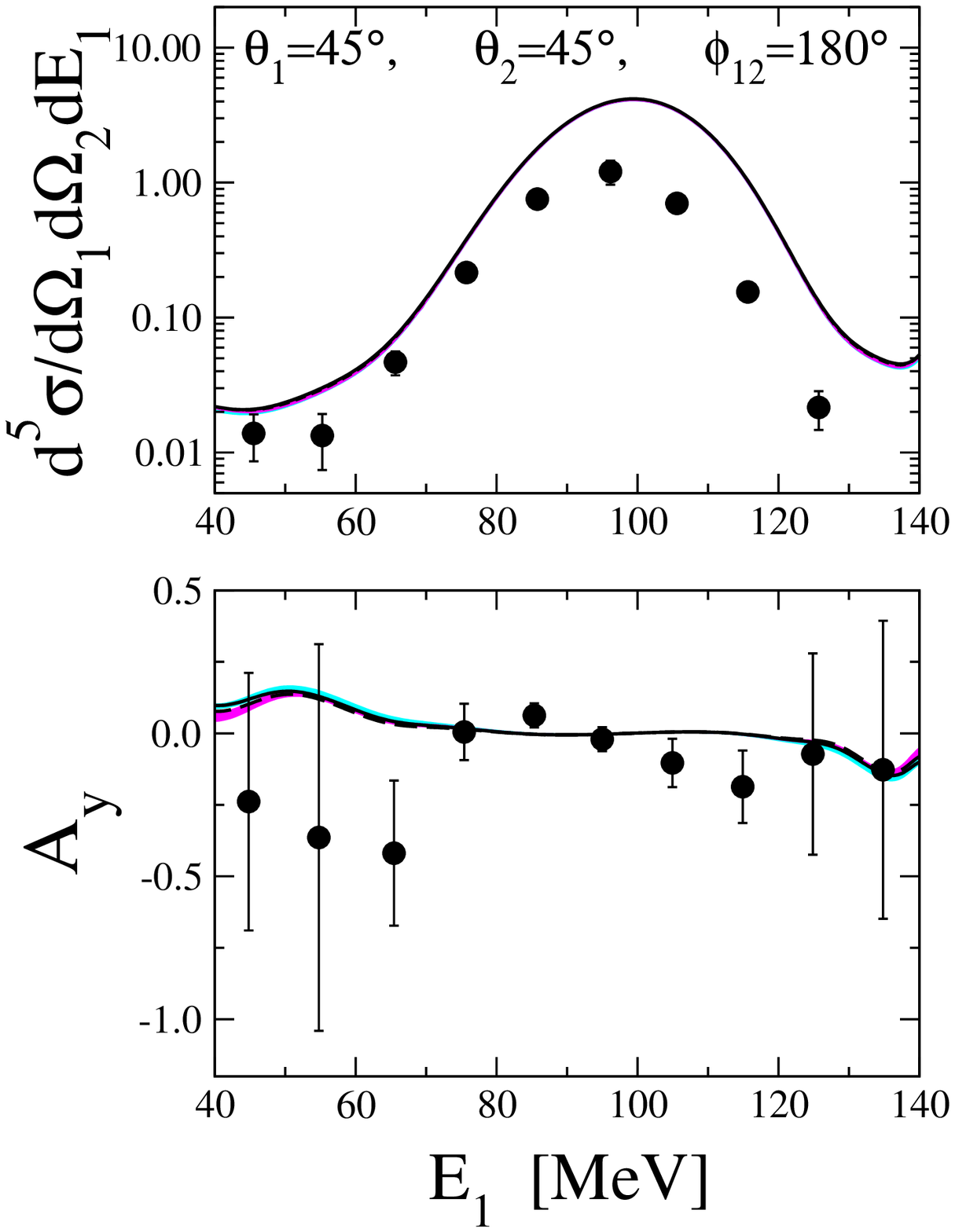}}}
  \caption[]{Nd breakup cross section in [mb MeV$^{-1}$sr$^{-2}$]
and nucleon analyzing power data at 200 MeV in comparison to theory. 
Bands and curves as in Fig.~\ref{fig:e13-sig-qfs}. The $pd$ data are from~\cite{pairsuwan:95}.
}
  \label{fig:e200-4}
\end{figure}

\begin{figure}[htbp]
\leftline{\mbox{\epsfxsize=17cm \epsffile{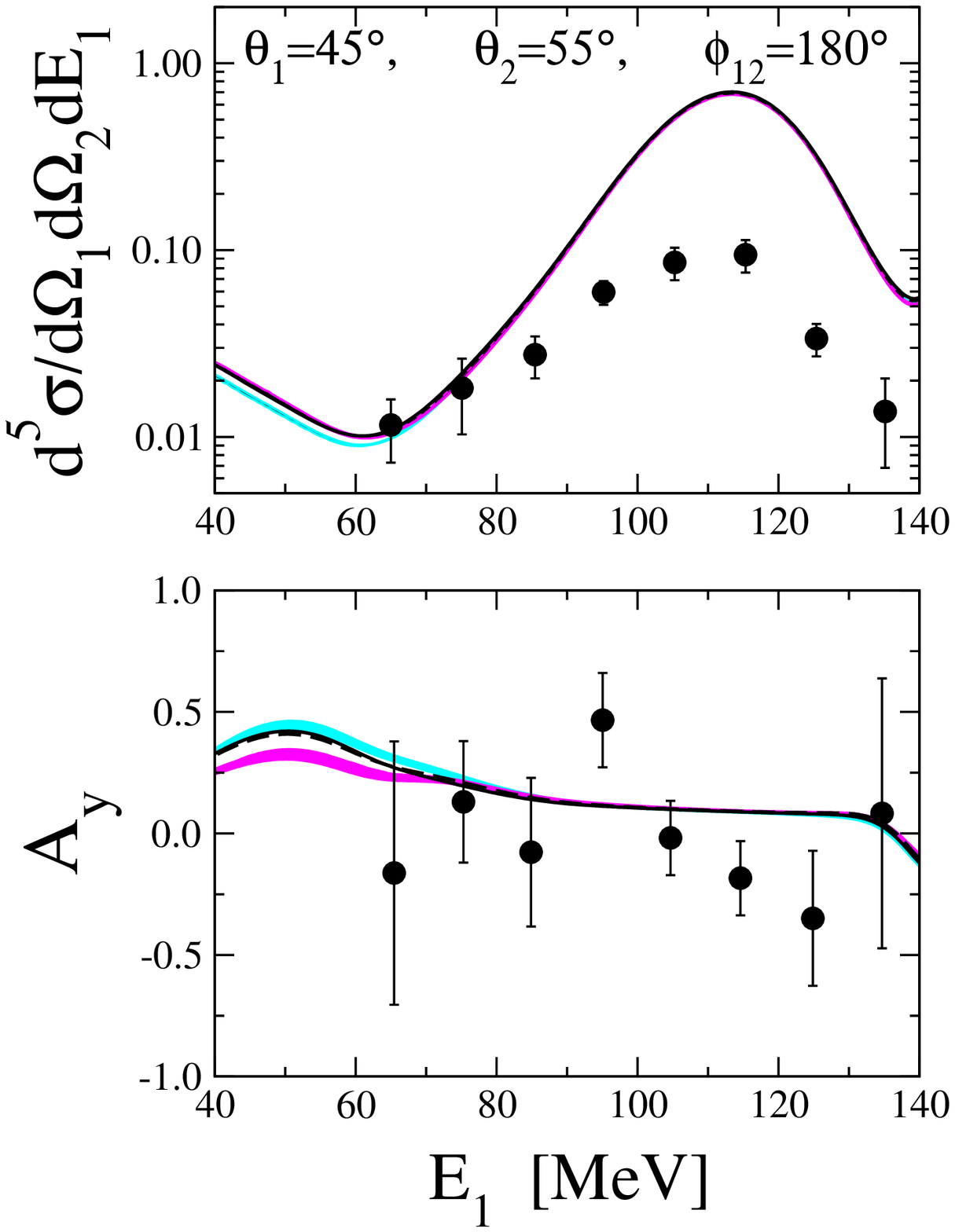}}}
  \caption[]{Nd breakup cross section in [mb MeV$^{-1}$sr$^{-2}$]
and nucleon analyzing power data at 200 MeV in comparison to theory. 
Bands and curves as in Fig.~\ref{fig:e13-sig-qfs}. The $pd$ data are from~\cite{pairsuwan:95}.
}
  \label{fig:e200-5}
\end{figure}

\begin{figure}[htbp]
\leftline{\mbox{\epsfxsize=17cm \epsffile{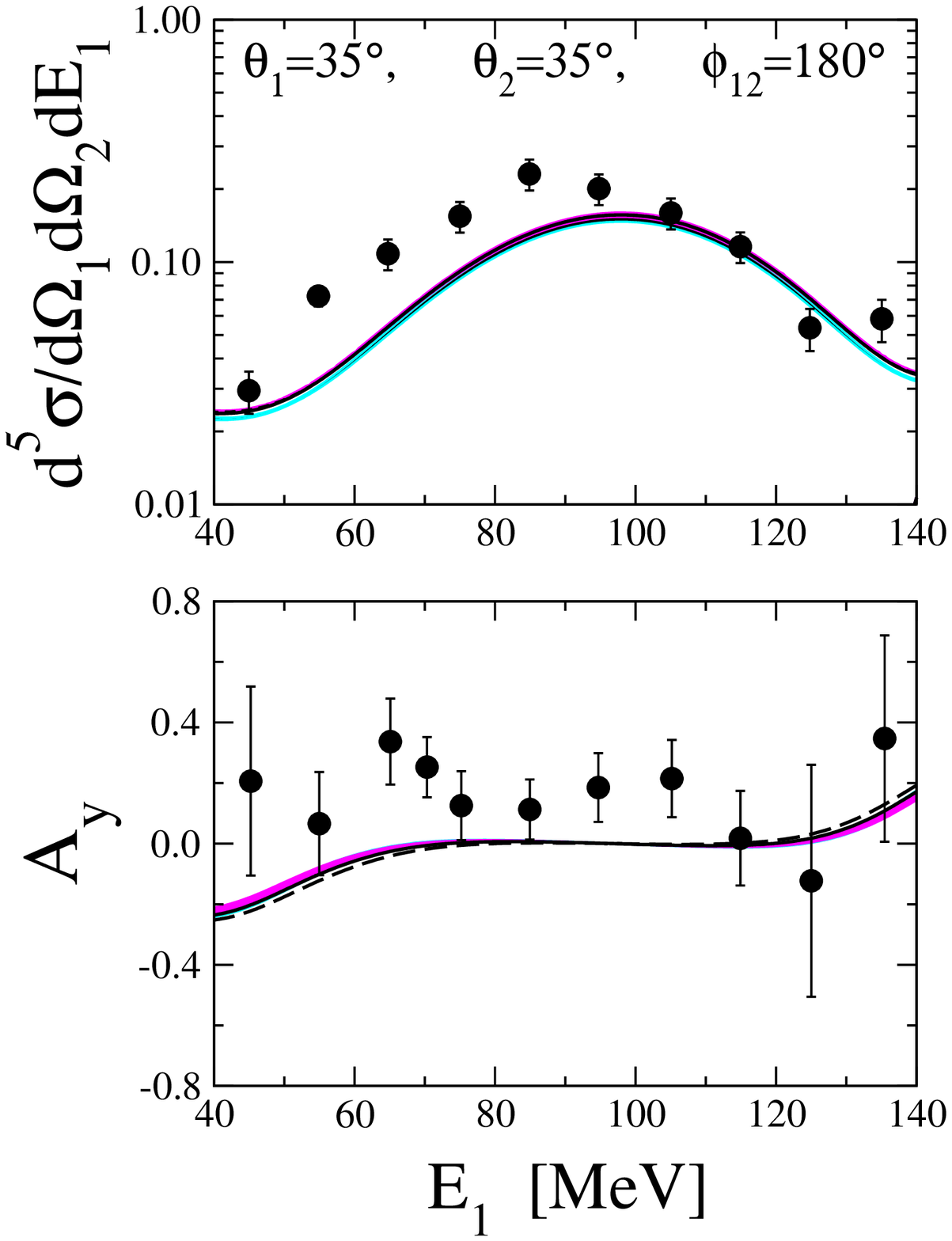}}}
  \caption[]{Nd breakup cross section in [mb MeV$^{-1}$sr$^{-2}$] 
and nucleon analyzing power data at 200 MeV in comparison to theory. 
Bands and curves as in Fig.~\ref{fig:e13-sig-qfs}. The $pd$ data are from~\cite{pairsuwan:95}.
}
  \label{fig:e200-6}
\end{figure}

\begin{figure}[htbp]
\leftline{\mbox{\epsfxsize=17cm \epsffile{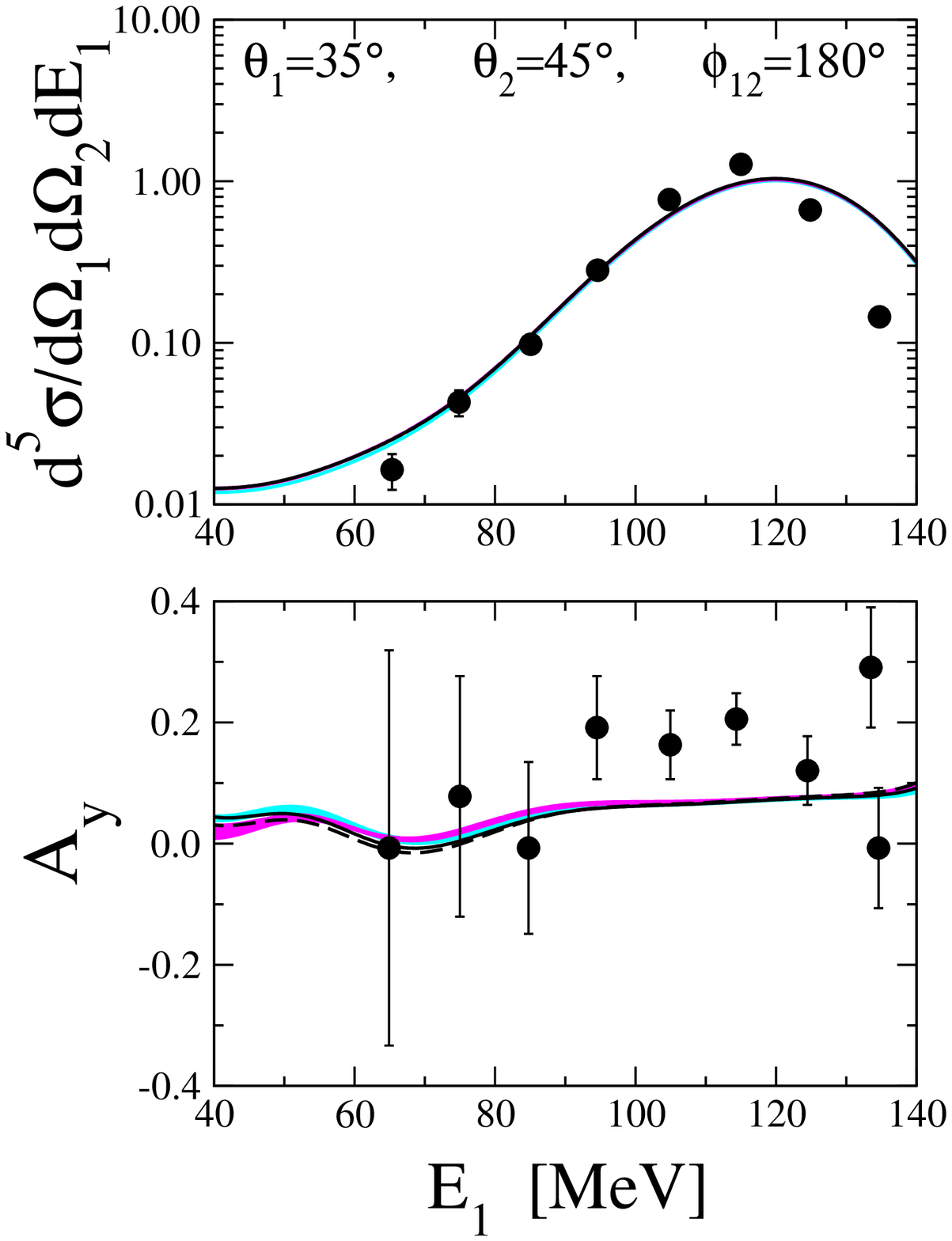}}}
  \caption[]{Nd breakup cross section in [mb MeV$^{-1}$sr$^{-2}$]
and nucleon analyzing power data at 200 MeV in comparison to theory. 
Bands and curves as in Fig.~\ref{fig:e13-sig-qfs}. The $pd$ data are from~\cite{pairsuwan:95}.
}
  \label{fig:e200-7}
\end{figure}

\begin{figure}[htbp]
\leftline{\mbox{\epsfxsize=17cm \epsffile{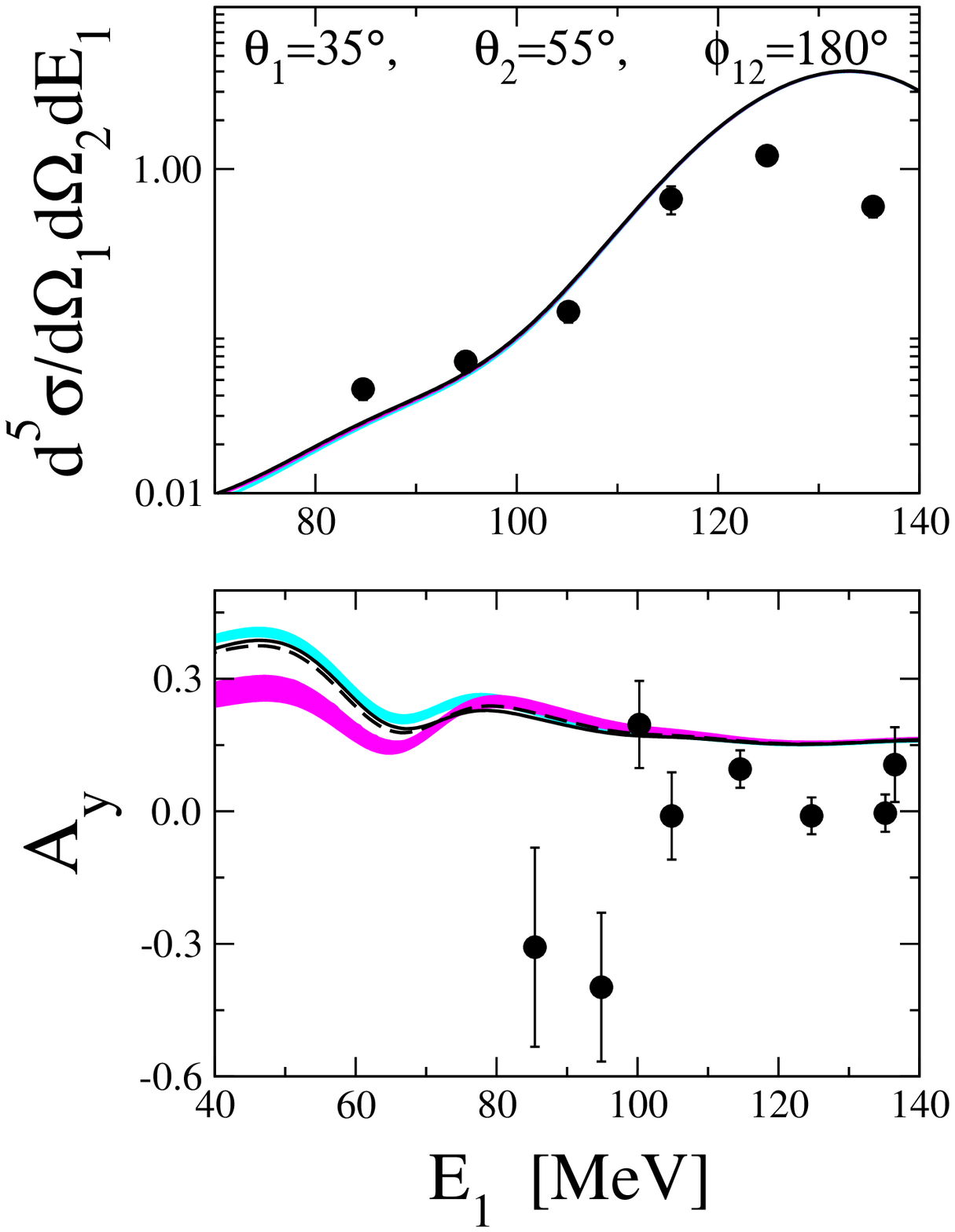}}}
  \caption[]{Nd breakup cross section in [mb MeV$^{-1}$sr$^{-2}$]
and nucleon analyzing power data at 200 MeV in comparison to theory. 
Bands and curves as in Fig.~\ref{fig:e13-sig-qfs}.
The $pd$ data are from~\cite{pairsuwan:95}.}
  \label{fig:e200-8}
\end{figure}

In Figs.~\ref{fig:e200-1}-\ref{fig:e200-8} we show a comparison 
of our theoretical predictions with the $pd$ 
data of~\cite{pairsuwan:95} for the cross 
section $\frac{d^5\sigma}{d\Omega_1 d\Omega_2 dE_1}$ and the 
nucleon analyzing
power $A_y$. 
For the cross section the two bands are 
very narrow and overlapping. Thus we predict 
practically no 3NF effects. It is no surprise, since most of the configurations 
are in the vicinity of QFS. The comparison with 
the data, however, shows striking 
disagreements in most cases. Though the 
shapes are generally quite well reproduced, the 
magnitudes are wrong. This is alarming, since 
the current nuclear forces fail strongly. 
Note, however, we have no estimate for relativistic effects, which at this high energy 
can contribute both kinematically and dynamically.

Also in case of $A_y$ the two bands are mostly 
rather narrow and overlapping. Since some of 
the data have large error bars, agreement 
or disagreement of theory and data is not clear.

We are not aware of other breakup data in that energy region.
The sensitive regions for 3NF effects are around 
$\theta_1 \sim $ 15~$^\circ \sim \theta_2$
and $ 0^\circ \le \phi_{12}  \le 20^\circ $
for the cross section and $\theta_1 \sim $ 100$^\circ$ , 
$\theta_2 \sim $ 30$^\circ$ (and vice versa)
and $\phi_{12}  \sim 180^\circ $ for $A_y$.

\section{Summary}
\label{secIV}

We compared modern NN force predictions alone and together with current 3NF 
models to a
selected set of Nd breakup cross sections and analyzing power data at 13, 65 
and 200 MeV.
Though in most cases the agreement was good, we also found cases with striking
discrepancies between theory and experiment. The discrepancies showed up in 
the SSS, QFS and
some unspecified geometries at low energies. Severe discrepancies are also 
present in
the cross sections at 200 MeV. In all those cases the 3NF effects predicted 
by the current
models are very small. At 200 MeV we can  not exclude that at least one reason for
the discrepancy might lie in the totally neglected relativistic effects. At 
the lower
energies $pp$ Coulomb effects, not included in our theoretical description, might 
also play a
role.
In case of the analyzing power $A_y$ we found some discrepancies at 65 MeV, 
which point to
deficiencies in the current nuclear force models. Some configurations with 
interesting
theoretical 3NF effects  at this energy could not be checked conclusively 
against
 experiment, since there
is a big scatter in the available data.

The experiments performed so far show that it is rather difficult to find by 
chance a
configuration with large 3NF effects. Therefore the breakup experiments 
should be
guided by theoretical predictions like the one in I. Also  the present day Nd 
breakup data set
is much  poorer than the elastic scattering one, what calls for more data. 
Especially
cross section and analyzing power measurements at higher energies in 
configurations where
large 3NF effects have been predicted are highly desirable.

\acknowledgements

This work was supported by
the Polish Committee for Scientific Research  (Grant No. 2P03B02818 
and 5P03B12320), 
 the Deutsche Forschungsgemeinschaft (J.G.), and the NSF (Grant No. PHY0070858).
One of us (W.G.) would like to thank the Foundation for Polish Science
for the financial support during his stay in Cracow.
The numerical calculations have been performed on the Cray T90 and T3E of the
NIC in J\"ulich, Germany.


\newpage 
\newpage 
\newpage 

\newpage 
\newpage 

\newpage 
\newpage 

\newpage 

\newpage 

\newpage 

\newpage 

\newpage 


\newpage 

\begin{figure}[htbp]
  \caption[]{Nd breakup cross section in [mb MeV$^{-1}$sr$^{-2}$] 
and nucleon analyzing power data at 65 MeV  
in comparison to theory. Symmetric space star (SSS) configuration is shown.
Bands and curves as in Fig.~\ref{fig:e13-sig-qfs}.
The $pd$ data are from~\cite{zejma:97}.}
 \label{fig:e65-sss}
\end{figure}

\begin{figure}[htbp]
  \caption[]{Nd breakup cross section in [mb MeV$^{-1}$sr$^{-2}$] 
and nucleon analyzing power data at 65 MeV in comparison to theory. 
Symmetric forward star (FPS) configuration is shown.
Bands and curves as in Fig.~\ref{fig:e13-sig-qfs}. The $pd$ data are from~\cite{zejma:97}.
}
  \label{fig:e65-fps}
\end{figure}

\begin{figure}[htbp]
  \caption[]{Nd breakup cross section in [mb MeV$^{-1}$sr$^{-2}$]
and nucleon analyzing power data at 65 MeV in comparison to theory. 
Backward plane star (BPS) configuration is shown.
Bands and curves as in Fig.~\ref{fig:e13-sig-qfs}. The $pd$ data are from~\cite{zejma:97}.
}
  \label{fig:e65-bps}
\end{figure}

\begin{figure}[htbp]
  \caption[]{Nd breakup cross section  in [mb MeV$^{-1}$sr$^{-2}$]
and nucleon analyzing power data at 65 MeV in comparison to theory. 
Quasi-free scattering (QFS) configuration is shown.
Bands and curves as in Fig.~\ref{fig:e13-sig-qfs}. The $pd$ data are from~\cite{zejma:97}.
}
  \label{fig:e65-qfs1}
\end{figure}

\begin{figure}[htbp]
  \caption[]{Nd breakup cross section in [mb MeV$^{-1}$sr$^{-2}$]
and nucleon analyzing power data at 65 MeV in comparison to theory. 
Quasi-free scattering (QFS) configuration is shown.
Bands and curves as in Fig.~\ref{fig:e13-sig-qfs}. The $pd$ data are from~\cite{zejma:97}.
}
  \label{fig:e65-qfs2}
\end{figure}

\begin{figure}[htbp]
\caption[]{Nd breakup cross section in [mb MeV$^{-1}$sr$^{-2}$] and nucleon analyzing power data
at 65 MeV in comparison to theory.
Collinear (COLL) configuration is shown.
Bands and curves as in Fig.~\ref{fig:e13-sig-qfs}.
The $pd$ data are from~\cite{allet:96}.}
\label{fig:e65-coll-116}
\end{figure}
\begin{figure}[htbp]
\caption[]{Nd breakup cross section in [mb MeV$^{-1}$sr$^{-2}$] and nucleon analyzing power data
at 65 MeV in comparison to theory.
Collinear (COLL) configuration is shown.
Bands and curves as in Fig.~\ref{fig:e13-sig-qfs}.
The $pd$ data are from~\cite{allet:96}.}
\label{fig:e65-coll-98}
\end{figure}

\begin{figure}[htbp]
\caption[]{Nd breakup cross section in [mb MeV$^{-1}$sr$^{-2}$] and nucleon analyzing power data
at 65 MeV in comparison to theory.
Collinear (COLL) configuration is shown.
Bands and curves as in Fig.~\ref{fig:e13-sig-qfs}.
The $pd$ data are from~\cite{allet:96}.}
\label{fig:e65-coll-75.6}
\end{figure}

\begin{figure}[htbp]
\caption[]{Nd breakup cross section in [mb MeV$^{-1}$sr$^{-2}$] and nucleon analyzing power data
at 65 MeV in comparison to theory.
Collinear  (COLL) configuration is shown.
Bands and curves as in Fig.~\ref{fig:e13-sig-qfs}.
The $pd$ data are from~\cite{allet:96}.}
\label{fig:e65-coll-59.5}
\end{figure}
\newpage


\newpage

\newpage

\newpage

\newpage

\newpage

\newpage

\newpage

\newpage

\thispagestyle{empty}

\end{document}